%% file: main.tex
\documentclass[acmtog]{acmart}
\acmSubmissionID{251}

\usepackage{booktabs} 
\usepackage{float}
\usepackage{stfloats}
\usepackage{animate}
\usepackage{siunitx}
\DeclareSIUnit{\nothing}{\relax}
\usepackage{amsmath}
\usepackage{esvect}
\DeclareMathAlphabet{\mathcal}{OMS}{cmsy}{m}{n}

\usepackage{url}                             
\usepackage[nameinlink,capitalise]{cleveref} 
\crefname{section}{Sec.}{Secs.}
\crefname{table}{Tab.}{Tabs.}
\usepackage{microtype}                       
\usepackage{wrapfig}                         
\usepackage{booktabs}                        
\usepackage{tabularx}
\usepackage{cancel}                          
\usepackage{mathtools}                       
\usepackage{xfrac}                           
\usepackage{nicefrac}                        
\usepackage[shortlabels,inline]{enumitem}    
\setlist[enumerate]{label=\arabic*.}
\usepackage{ellipsis}                        
\usepackage{xspace}                          
\usepackage{bm}                              
\usepackage{placeins}
\usepackage[detect-none]{siunitx}
\usepackage{custom_command}
\usepackage{algorithm}
\usepackage[noend]{algpseudocode}
\usepackage{soul}
\usepackage{listings}

\acmJournal{TOG}

\begin{document}
\setcopyright{cc}
\setcctype{by}
\acmJournal{TOG}
\acmYear{2026} \acmVolume{45} \acmNumber{4} \acmArticle{89}
\acmMonth{7} \acmDOI{10.1145/3811299}

\title{ToF ReSTIR: Time-of-Flight Rendering with Spatio-temporal Reservoir Resampling}


\author{Juhyeon Kim}
\affiliation{%
  \institution{Dartmouth College}
  \country{USA}
}
\email{juhyeon.kim.gr@dartmouth.edu}
\orcid{0000-0002-6218-3426}

\author{Wojciech Jarosz}
\affiliation{%
 \institution{Dartmouth College}
  \country{USA}
}
\email{wojciech.k.jarosz@dartmouth.edu}
\orcid{0000-0002-1652-0954}

\author{Adithya Pediredla}
\affiliation{%
  \institution{Dartmouth College}
  \country{USA}
}
\email{adithya.k.pediredla@dartmouth.edu}
\orcid{0000-0002-6623-020X}

\renewcommand\shortauthors{Kim et al.}

\input{sections/teaser}
\input{sections/00_abstract_v1.0}
%
%
\maketitle

\input{sections/01_introduction_v1.5}
\input{sections/02_related_works_v1.6}
\input{sections/03_background_v1.3}
\input{sections/04_method_v1.2}

\input{sections/05_implementation_v1.1}

\input{sections/06_results_v1.2}
\input{sections/07_applications_v1.0}
\input{sections/08_conclusion_v1.1}

\bibliographystyle{ACM-Reference-Format}
\bibliography{bib/strings-full, bib/rendering-bibtex, bib/additional, bib/FMCW_References_Barber}

\end{document}

%% file: sections/teaser.tex
\begin{teaserfigure}
\centering
    \includegraphics[width=\linewidth]{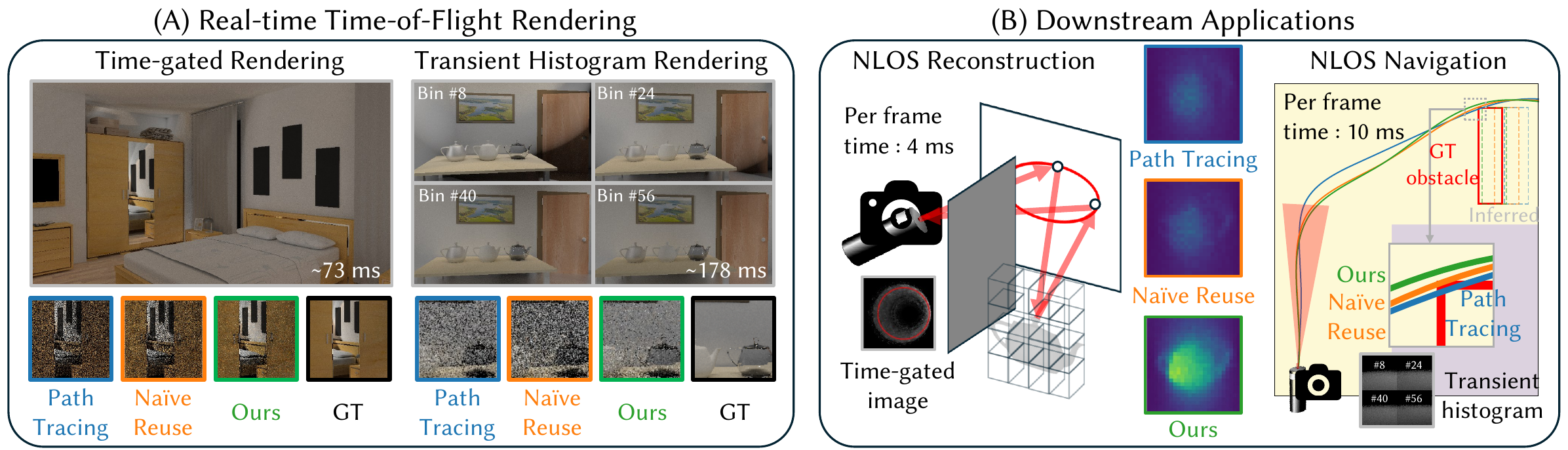}
  \caption{
  We propose a real-time Time-of-Flight (ToF) rendering method inspired by ReSTIR, a state-of-the-art technique for steady-state real-time rendering, augmented with a novel path-length-aware shift mapping.
  (A) Compared to conventional path tracing and naive path reuse adapted from steady-state ReSTIR~\cite{Lin:2022:Generalized}, our method produces improved visual results for both time-gated image rendering and transient histogram rendering in dynamic scenes at interactive frame rates.
  (B) We further demonstrate two interactive downstream applications enabled by our renderer.
  For non-line-of-sight (NLOS) reconstruction, we recover hidden voxel geometry using temporally focused imaging~\cite{Pediredla:2019:SNLOS}, where time-gated image simulation with our method reliably reconstructs the shape of the teapot.
  We also simulate NLOS navigation tasks~\cite{young2025enhancing}, where the reduced noise in our rendered transient histograms enables robust neural network inference and reliable path planning, whereas baseline methods fail to produce safe paths due to noise.
  }
  \label{fig:teaser}
\end{teaserfigure}

%% file: sections/00_abstract_v1.0.tex
\begin{abstract}
We present a novel spatio-temporal reuse framework for time-resolved light transport, enabling efficient Monte Carlo rendering of time-of-flight (ToF) phenomena such as time-gated imaging and transient light capture.
Existing ToF rendering methods are computationally expensive, scale poorly to complex dynamic scenes, and are therefore unsuitable for applications with strict latency constraints.
To address this limitation, we draw inspiration from \emph{ReSTIR}, a reuse-based technique for steady-state real-time rendering, and adapt its core principles to interactive-rate ToF simulation.
However, naively applying existing ReSTIR methods to ToF rendering leads to severe inefficiency, as reused paths frequently violate optical path-length constraints and thus contribute little or no signal.
We overcome this challenge by introducing a path reuse formulation that explicitly enforces physically valid optical path lengths.
The key idea is \emph{path-length-aware shift mapping}, a geometric transformation based on Newton’s method that adjusts reused light paths to satisfy temporal gating constraints, inspired by specular manifold exploration in steady-state caustics rendering.
The resulting framework substantially improves the efficiency of ToF rendering across a wide range of scenarios, including complex scenes with glossy or specular materials and dynamic motion.
Our method supports both time-gated and transient rendering at interactive frame rates, enabling simulation under practical latency constraints.
We demonstrate the effectiveness of our approach through two downstream applications, including shape reconstruction and navigation.

\end{abstract}

%% file: sections/01_introduction_v1.5.tex
\section{Introduction}

Light travels fast, but not infinitely fast—and even small delays in a photon’s arrival carry rich information about the physical world. 
Time-of-flight (ToF) imaging leverages this principle by emitting controlled illumination and measuring when photons return to the sensor after interacting with the scene. 
These time-resolved measurements capture both direct reflections from visible surfaces and multiply scattered transport, often originating from partially or fully hidden objects. 
As a result, ToF imaging enables advanced sensing capabilities—such as non-line-of-sight reconstruction and material characterization in challenging environments—that are inaccessible to conventional steady-state intensity cameras.

Time-of-flight rendering provides a physically grounded way for simulating measurements from time-resolved imaging systems and plays an important role in understanding, validating, and developing modern ToF sensors.
Such simulations support offline analysis and algorithm development, and enable the generation of ground-truth datasets that may be difficult or impractical to acquire experimentally.
Beyond these offline uses, improved rendering performance enables ToF-based systems to be evaluated in interactive and decision-driven settings.
For example, applications such as real-time reconstruction, navigation, and closed-loop perception require simulation to operate within strict latency budgets in order to be applicable.
Efficient and physically accurate ToF rendering therefore, can extend the scope of time-resolved simulation from offline analysis to interactive sensing and perception tasks.

Despite recent progress, existing ToF rendering methods face a fundamental trade-off. 
Physically accurate approaches~\cite{Jarabo:2014:Framework, Marco:2019:Progressive, Pediredla:2019:Ellipsoidal, Liu:2022:Temporally} are often computationally expensive and limited to offline use, while faster methods~\cite{Iseringhausen:2020:Nonlineofsight, Klein:2016:Tracking} rely on simplified assumptions such as reduced geometry, materials, or light transport. 
As a result, current techniques struggle to achieve both efficiency and physical generality in dynamic, large-scale scenes.

\begin{figure}
\includegraphics[width=\linewidth]{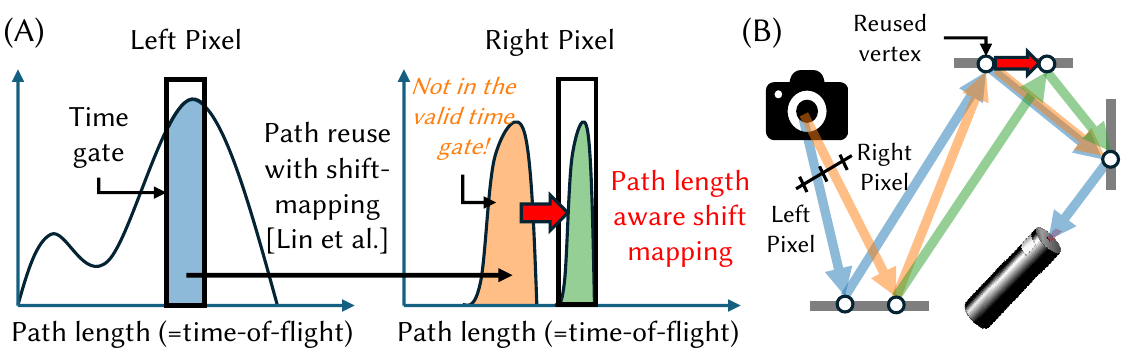}
\caption{
Overview of proposed path-length-aware reuse strategies. 
(A) Naive path reuse from conventional steady-state ReSTIR (e.g., hybrid shift-mapping by \citet{Lin:2022:Generalized}) ends up going outside of the valid time gate.
To address this issue, we introduce path-length-aware shift mapping, illustrated by the red arrow.
(B) Illustration of a base path in the left pixel (blue), the naively reused path in the right pixel (orange), and the corrected path produced by our method (green), which satisfies the temporal constraint.
}
\label{figure:overview_figure}
\end{figure}

Motivated by these limitations, it is natural to ask whether techniques developed for real-time physically based steady-state rendering (e.g., games or commercial rendering) can be adapted to support interactive and scalable ToF simulation.
In particular, reuse-based methods from Monte Carlo path tracing provide a promising starting point, as they aim to reduce computational cost while preserving physical correctness.
Reservoir-based spatio-temporal resampling, so-called \emph{ReSTIR}~\cite{Bitterli:2020:Spatiotemporal} and related approaches~\cite{Ouyang:2021:ReSTIR, Lin:2022:Generalized} are representative methods in this direction, demonstrating how aggressive spatio-temporal reuse can scale path tracing to dynamic and interactive settings.
However, these methods are fundamentally designed for steady-state rendering and do not account for optical path length.
When applied directly to ToF rendering, reused paths frequently violate temporal constraints, 
leading to invalid contributions (\cref{figure:overview_figure}).

To address these challenges, we introduce a path-length-aware spatio-temporal reuse framework for ToF rendering.
The key idea is \emph{path-length-aware shift mapping}, which geometrically adjusts a reused path so that its optical path length matches the active time gate (red arrow in \cref{figure:overview_figure}).
We formulate this adjustment as a constrained root-finding problem and solve it using Newton’s method.
Our approach is inspired by manifold exploration techniques for specular transport~\cite{Jakob:2012:Manifold, Lehtinen:2013:Gradientdomain, hong2025sample}, but is adapted to enforce path-length constraints specific to ToF rendering.
By incorporating path-length awareness into ReSTIR-style reuse, our method enables scalable and efficient simulation of ToF sensors in complex, dynamic environments.
Our reuse strategy supports both per-frame image and transient histogram rendering, and we demonstrate its effectiveness in two downstream NLOS applications: shape reconstruction and inference-based navigation (\cref{fig:teaser}).

The code and data are available on the project page\footnote{\url{https://juhyeonkim95.github.io/project-pages/tof_restir}}.

%% file: sections/02_related_works_v1.6.tex
\section{Related Works}
\subsection{Time-of-Flight Imaging}
Here, we review representative ToF imaging systems that acquire either transient measurements—sequences of time-resolved images—or time-gated measurements, which capture a single image within a narrow temporal window.
\Citet{Jarabo:2017:Recent} provide a more comprehensive review.

One common approach employs high-speed optical gating to selectively detect photons that arrive within short temporal windows after pulsed illumination. 
Intensified CCD/CMOS cameras achieve this using a fast-gating image intensifier, allowing selective capture of early or late light within sub-nanosecond windows~\cite{cester2019time}. By shifting these gates over repeated acquisitions, a full transient profile can be reconstructed. 
Streak cameras achieve even finer temporal resolution by mapping photon arrival times to spatial deflections, enabling picosecond-scale measurements~\cite{Velten:2012:Recovering, Velten:2013:Femtophotography}.
Single-photon avalanche diode (SPAD) sensors have emerged as compact, scalable detectors operating at the level of individual photons~\cite{zappa2007principles}. In time-correlated single-photon counting (TCSPC) mode, SPAD arrays time-stamp photon arrivals to build transient histograms at each pixel~\cite{Gariepy:2015:Singlephoton,henderson2019192}. 
Continuous-wave ToF cameras are widely used for conventional range imaging, but transient information can also be recovered by multi-frequency phase measurements~\cite{Heide:2013:Lowbudget, Peters:2015:Solving}. 
There are also interferometric approaches that measure optical phase with micron-scale path-length sensitivity~\cite{Gkioulekas:2015:Micronscale, Kadambi:2016:Macroscopic}.


ToF imaging has become an important tool in computational imaging and robotics.
Time-resolved transient measurements enable advanced tasks such as non-line-of-sight (NLOS) reconstruction~\cite{Laurenzis:2014:Nonlineofsight, Buttafava:2015:Nonlineofsight, Kirmani:2011:Looking, Velten:2012:Recovering, somasundaram2023role, tsai2017geometry}, separation of direct and indirect transport~\cite{Wu:2014:Decomposing, OToole:2014:Temporal}, and robust perception for autonomous navigation~\cite{young2025enhancing}.
Time-gated imaging has a narrower range of applications but remains effective for depth-selective sensing in autonomous vehicle~\cite{grauer2015active} or military applications~\cite{baker2004low}, and some NLOS applications~\cite{Pediredla:2019:SNLOS}.
Additionally, transient capture realized by shifting gating delays behaves as time-gated imaging for dynamic scenes, since each gate samples a slightly different configuration.


\subsection{Time-of-Flight Rendering}
Time-of-flight (ToF) rendering simulates measurements from transient and time-gated sensors by explicitly incorporating optical path length into light transport. Following~\citet{Pediredla:2019:Ellipsoidal, Liu:2022:Temporally}, we use this term to refer to both transient rendering (time-resolved image sequences, i.e., histogram) and time-gated rendering (single images within a temporal window).

A key challenge shared by both ToF rendering tasks arises from the presence of a narrow temporal manifold induced by a time gate or a temporal bin.
As a result, naive steady-state rendering algorithms tend to fail, since most sampled light paths fall outside the valid temporal range and contribute nothing.
For transient rendering under a static scene, this can be partially mitigated by sharing samples across all temporal bins
\footnote{
This is called \emph{path reuse} in \citet{Jarabo:2014:Framework}. This keeps the sampled path unchanged over temporal bins, different from path reuse in ReSTIR, which is in a more aggressive form that can even deform the path (e.g., reconnect to a new vertex) for reuse.
}, 
where each sampled path is accumulated into its corresponding time bin~\cite{Jarabo:2012:Femtophotography, Marco:2013:Transient, Ament:2014:Refractive, Adam:2016:Bayesian}, performing histogram density estimation.
To obtain accurate transient profiles, a small bin width is required, but this often leads to noisy histograms.
To address this issue, \citet{Jarabo:2014:Framework} introduced progressive kernel-density estimation in the temporal domain, trading variance for bias.
This was later extended to progressive spatio-temporal density estimation using photon beams~\cite{Jarosz:2011:Progressive} under participating media~\cite{Marco:2019:Progressive}.
An alternative strategy of the temporal blurring is directly determining the time stamp and only sampling light paths that are valid at that specific time stamp.
This is essential for time-gated rendering and also beneficial for transient rendering, as uniformly distributed samples are helpful for kernel density estimation.
Prior work has explored path-length-constrained sampling in participating media~\cite{Jarabo:2014:Framework} and surface transport via ellipsoidal path connection~\cite{Pediredla:2019:Ellipsoidal}, while more recent methods slice photon primitives into temporal wavefronts for efficient filtering~\cite{Liu:2022:Temporally}.

Beyond forward rendering, differentiable ToF formulations~\cite{Yi:2021:Differentiable, Wu:2021:Differentiable} and extensions to Doppler effects~\cite{kim2023doppler, kim2025monte} have also been explored.



From a performance perspective, none of the previous ToF rendering methods targeted interactive rate rendering for complex scenes.
Several fast radiosity-based approaches on GPU have been proposed to accelerate simulation by modeling objects as Lambertian patches~\cite{Klein:2016:Tracking} or interpolating radiance over mesh triangle vertices~\cite{Iseringhausen:2020:Nonlineofsight}.
However, these methods are typically limited to two-bounce light transport and often require heavy mesh preprocessing.
Recently, some GPU-based transient path tracers have also been introduced, but they are still not designed for interactive purposes~\cite{Royo:2022:Nonlineofsight, royo2025mitransient}.

Overall, while physically accurate ToF rendering techniques exist, they remain computationally expensive and difficult to apply to large-scale, dynamic environments at interactive rates on a GPU.
This motivates the development of scalable ToF rendering approaches inspired by modern real-time steady-state rendering techniques, while explicitly accounting for temporal constraints.


\subsection{Spatio-Temporal Path Reuse}
Reservoir-based spatiotemporal reuse, or ReSTIR, has recently transformed steady-state real-time Monte Carlo rendering~\cite{Bitterli:2020:Spatiotemporal}.
ReSTIR leverages an importance-resampling strategy~\cite{Talbot:2005:Importance} to reuse light samples across neighboring pixels and successive frames, achieving high-quality results with only a few samples per pixel.
Follow-up works have extended these principles to global illumination~\cite{Ouyang:2021:ReSTIR}, volumetric transport~\cite{Lin:2021:Fast}, general multi-bounce path tracing~\cite{Lin:2022:Generalized}, depth of field and antialiasing~\cite{zhang2024area}, motion blur~\cite{liu2025reservoir}, BDPT~\cite{hedstrom2025restir}, path guiding~\cite{zeng2025restir}, and specular manifold sampling~\cite{hong2025sample}.
These methods excel in scenarios where spatial and temporal coherence can be exploited, achieving high-quality results with real-time performance even in complex scenes.
However, they assume a steady-state setting and are agnostic to optical path length.
As a result, when directly applied to ToF sensors, reused paths often become temporally invalid and fail to yield meaningful variance reduction.

\subsection{Sampling on Specular Manifold}
Sampling a narrow temporal manifold is closely related to sampling specular manifolds in steady-state rendering.
A representative example is \emph{manifold walk}~\cite{Jakob:2012:Manifold}, which proposes finding valid specular paths from initially invalid mutated MCMC paths using Newton’s iteration.
This idea was later extended to half-vector space light transport~\cite{Kaplanyan:2014:Naturalconstraint} and manifold next-event estimation~\cite{Hanika:2015:Improved, Zeltner:2020:Specular}.
Several follow-up works further improved specular manifold sampling through techniques such as manifold path guiding~\cite{Li:2022:Unbiased, fan2023manifold} and spatio-temporal reuse~\cite{xu2023efficient}.
More recently, \citet{hong2025sample} adopted sample-space partitioning to improve specular manifold sampling and demonstrated its effectiveness when implemented within a ReSTIR framework.
Our temporal manifold path reuse is closely related to these specular manifold methods, but differs in the underlying degrees of freedom of the constraint manifold.
We address this difference by introducing an additional gauge condition to uniquely determine valid solutions.

%% file: sections/03_background_v1.3.tex
\section{Background}

\subsection{Time-of-Flight Rendering}
The path integral formulation of conventional steady-state Monte Carlo rendering is given by~\cite{Veach:1997:Robust}
\begin{equation}
\label{eq:path_integral}
    I = \int_\pathspace f(\xbar) \dxbar
\end{equation}
where $I$ denotes the measurement at a pixel, $\pathspace$ is the set of all possible light paths $\xbar$, consisting of vertices $\mathbf{x}_0, ..., \mathbf{x}_K$, $f(\xbar)$ is the path throughput and $\mu$ is its Lebesgue measure.
For ToF rendering, the path integral formulation is slightly modified as~\cite{Jarabo:2014:Framework, Pediredla:2019:Ellipsoidal}
\begin{equation}
\label{eq:tof_path_integral}
    I(\tau) = \int_\pathspace W_\tau\cP{\pathLength{\xbar}} f(\xbar) \dxbar
\end{equation}
where $W_\tau$ is \emph{path-length importance} function at time $\tau$, and $\pathLength{\xbar}$ is the optical path length defined as $\sum_{k=0}^{K-1}\eta_{k}\len{\mathbf{x}_{k+1}-\mathbf{x}_{k}}$ where $\eta_{k}$ is the refractive index.
For simplicity, we set the speed of light as $c=1$ throughout the paper.
For transient histogram rendering, we record a sequence of images for different $\tau\in\{\tau_1, ..., \tau_B\}$ where $B$ is the number of histogram bins.


\subsection{Resampled Importance Sampling and ReSTIR}
We next review the theoretical foundations of ReSTIR for evaluating the path integral in \cref{eq:path_integral}.
This summary is based on~\citet{Lin:2022:Generalized, Sawhney:2022:Decorrelating}, and we recommend readers to read the original papers for additional details.

\subsubsection{Basic RIS (Same Domain, Same PDF)}
Resampled Importance Sampling (RIS)~\cite{Talbot:2005:Importance} aims to estimate the integral of a function $f$ using a non-negative unnormalized target function $\hat{p}(x)$ in an unbiased manner.
In its basic form, RIS begins by generating $M$ independent candidate samples $\{x_1,\ldots,x_M\}$ from a domain $\Omega$ according to a pdf $p$ that is easy to sample.
From these candidates, a single representative sample $y=x_s$ is selected with a probability proportional to $w_s / \sum_{i=1}^M w_i$, where a resampling weight $w_i$ is
\begin{equation}
    \label{eq:ris_same_domain_same_pdf}
    w_i = \frac{1}{M} \hat{p}(x_i) W_i,
    \hspace{4pt}
    W_i := \frac{1}{p(x_i)} .
\end{equation}
The factor $W_i$ is known as the \emph{unbiased contribution weight} (UCW).
The probability density function of the selected sample $y$, denoted by $p_Y(y)$, which is required for traditional Monte Carlo integration, is generally unavailable in a closed form.
Instead, RIS computes the UCW of the winner as
\begin{equation}
W_y
:=
\frac{1}{\hat{p}(y)}
\sum_{i=1}^M w_i ,
\end{equation}
which can be used as an effective substitute for $1/p_Y(y)$~\cite{Bitterli:2020:Spatiotemporal}.
This yields the RIS estimator for an integrand $f$,
\begin{equation}
\cA{I}_{\text{RIS}} = f(y) W_y .
\end{equation}
Provided that $p_Y(y)$ is non-zero wherever $f(y) > 0$ (i.e., over the support of $f$), the estimator remains unbiased.
A typical choice is to set $\hat{p}(x) = f(x)$, so that RIS asymptotically achieves an ideal importance sampling or zero-variance estimator.

\subsubsection{Combining with MIS  (Same Domain, Different PDF)}
RIS can be extended to incorporate multiple importance sampling (MIS) when candidate samples are generated using different sampling strategies (e.g., BSDF, light sampling), while still operating over the same domain $\Omega$.
In this setting, each candidate sample $x_i$ is drawn from its own proposal distribution $p_i$.
To combine these samples consistently, a resampling MIS scheme~\cite{Talbot:2005:Importancea} is applied, analogous to the balance heuristic~\cite{Veach:1997:Robust}.
Specifically, each sample is assigned a mixing weight given by
\begin{equation}
m_i(x_i)
=
\frac{p_i(x_i)}{\sum_{j=1}^{M} p_j(x_i)}.
\end{equation}
These mixing weights replace the uniform factor $1/M$ used in the i.i.d. case (\cref{eq:ris_same_domain_same_pdf}) and lead to the generalized resampling weight
\begin{equation}
w_i = m_i(x_i)\hat{p}(x_i)W_i,
\hspace{4pt}
W_i = \frac{1}{p_i(x_i)} .
\end{equation}
As long as the collection of proposal distributions ${p_i}$ jointly covers the support of $\hat{p}$, the resulting RIS estimator remains unbiased.


\subsubsection{Generalized RIS (Different Domain)}
Resampling candidates may be drawn from different target domains (reused across different pixels or time steps), which makes the previous assumption that all samples share the domain $\Omega$ invalid.
Generalized RIS~\cite{Lin:2022:Generalized} addresses this limitation by allowing candidate samples $x_i$ to be drawn from different domains $\Omega_i$, each associated with its own target function $\hat{p}_i$.
The key idea that enables resampling across different domains is \emph{shift mapping}, which defines a bijective transformation between domains.
Note that shift mapping does not always need to be successful, as we can reject the reuse if shift mapping fails.
Given a candidate sample $x_i \in \Omega_i$, a shift map $S_i : \Omega_i \rightarrow \Omega$ maps it to a corresponding sample $y_i = S_i(x_i) \in \Omega$.
Under this formulation, the resampling weight is given by
\begin{equation}
\label{eq:GRIS_resampling_weight}
w_i = m_i(y_i)\hat{p}(y_i)W_i
\cdot
\cB{\frac{\partial S_i}{\partial x_i}}.
\end{equation}
To compute the MIS weights $m_i$, \citet{Lin:2022:Generalized} introduce a generalized balance heuristic given by
\begin{equation}
\label{eq:GRIS_MIS_weight}
 m_i(y_i) = \frac{M_i \hat{p}_{\leftarrow i} (y_i)}{\sum^{M}_{j=1} M_j \hat{p}_{\leftarrow j}(y_j)}
\end{equation}
where $M_i$ denotes a confidence weight, and the term ``$\hat{p}$ from $i$'' is defined as $\hat{p}_{\leftarrow i} (y_i) = \hat{p}_i(x_i)\cB{\partial S_i^{-1}/\partial y_i}$ or 0 if $S_i^{-1}$ is not valid.
This formulation remains applicable even when the $x_i$ originate from RIS steps, such as merging winner samples from neighboring pixels, in which case the proposal density $p_i$ may not be explicitly available.

Selecting a good shift-mapping $S$ is critical, since a sample that is important in one domain may contribute little or nothing after being mapped to another domain where the integrand changes significantly.
A variety of shift-mapping strategies have been proposed in prior works.
Path reconnection~\cite{Lehtinen:2013:Gradientdomain, Ouyang:2021:ReSTIR} reconnects base and offset paths at the earliest compatible surface interaction and performs well for moderately rough materials.
Alternative strategies include half-vector copying~\cite{kettunen2015gradient} and random replay~\cite{Hua:2019:Survey}, which are particularly effective for specular transport.
To handle a broad class of materials robustly, \citet{Lin:2022:Generalized} propose a hybrid shift-mapping approach that mixes path reconnection and random replay (\cref{figure:path_length_shiftmap_overview}).

We note that existing shift-mapping techniques are designed for steady-state rendering, which tries to match $f(\xbar)$ after the reuse.
In ToF rendering, however, the integrand additionally includes a path-length-dependent term $W_\tau(\pathLength{\xbar})$, which introduces new challenges that are not addressed by prior methods.

\subsubsection{ReSTIR}
ReSTIR builds upon previous RIS-based resampling strategies to enable sample sharing across pixels, both spatially within a single frame and temporally across consecutive frames.
Rather than explicitly storing all $M$ samples used in RIS, ReSTIR employs Weighted Reservoir Sampling (WRS)~\cite{Chao:1982:General} to perform resampling in an incremental manner.
This approach uses a compact data structure, referred to as a \emph{reservoir}, which stores only the current winner sample and its associated contribution weight.
A typical ReSTIR pipeline proceeds through the following stages:
\begin{enumerate}[label=(\arabic*)]
    \item Initial sampling: Generate multiple initial candidates and apply RIS to select a single representative sample, which is then stored in the reservoir.
    \item Temporal resampling: Reuse samples across consecutive frames, often using motion vectors.
    \item Spatial resampling: Select reservoirs from neighboring pixels and merge them into the current reservoir.
    \item Final shading: Compute the final pixel estimate by evaluating $f(y)W_y$ using the reservoir sample.
\end{enumerate}
In the following sections, we describe how this ReSTIR framework can be extended to support ToF rendering, where path-length constraints must be explicitly enforced.









%% file: sections/04_method_v1.2.tex
\section{Path Length Aware Shift Mapping}
In this section, we introduce path-length-aware shift mapping strategies that address the path-length mismatch arising from naive path reuse.
We further show efficient initial RIS sample generation from previously proposed ellipsoidal path connections~\cite{Pediredla:2019:Ellipsoidal}, or \emph{shrinking} from samples with wider gate, which also exploits path-length-aware shift mapping.
Finally, we show how our method extends to transient and Doppler rendering scenarios.

\begin{figure}
    \centering
    \includegraphics[width=\linewidth]{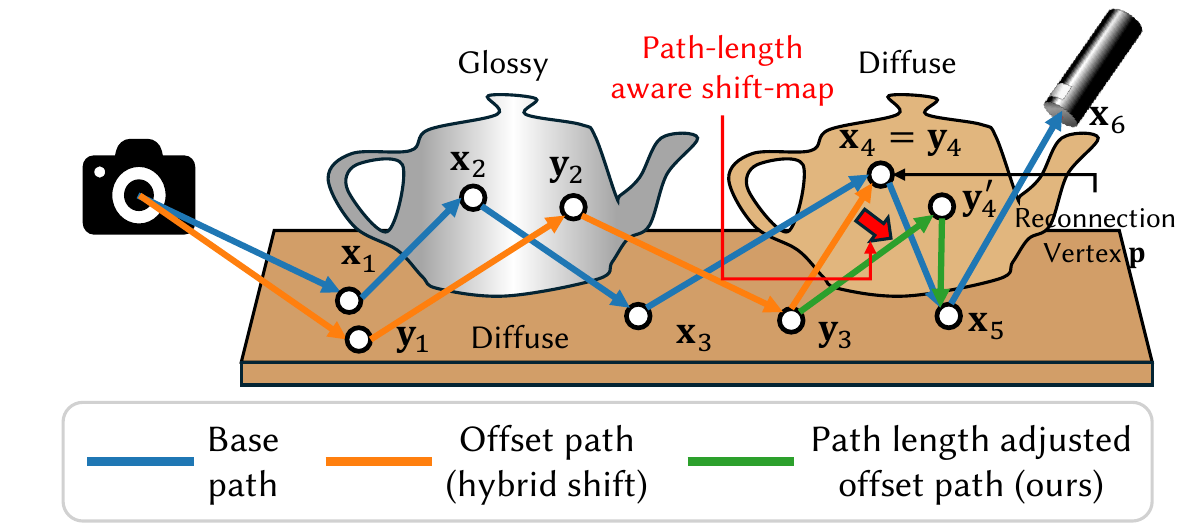}
    \caption[Path-length-aware shift mapping overview]{
    Baseline shift mapping (hybrid shift mapping~\protect\cite{Lin:2022:Generalized}) transforms the base path \ensuremath{\xbar} (blue) into an offset path \ensuremath{\ybar} (orange), with the reconnection vertex \ensuremath{\pbf} at \ensuremath{\ybf_4 (=\xbf_4)}.
    We further perturb the offset path \ensuremath{\ybar} to \ensuremath{\ybar'} (green) to satisfy the path-length constraint, \ensuremath{\pathLength{\ybar'} = \pathLength{\xbar} + \pathLengthDelta}.
    Rather than perturbing the entire offset path, we apply path-length-aware shift mapping only at the reconnection vertex, similar to \protect\citet{Sawhney:2022:Decorrelating}.
    (This figure is adapted from Fig.~6 of that paper to highlight the similarity.)
    }
    \label{figure:path_length_shiftmap_overview}
\end{figure}


\subsection{Offset Path Further Perturbation}
We begin with a baseline shift-mapping algorithm which transforms a \emph{base path} $\xbar$ into a shift-mapped \emph{offset path} $\ybar$.
Specifically, we use hybrid shift-mapping ~\cite{Lin:2022:Generalized} that uses random seed replay and reconnect at the first consecutive diffuse vertex (so called \emph{reconnection vertex} $\pbf$, corresponding to $\ybf_4$ in \cref{figure:path_length_shiftmap_overview}).
Now, the goal of path-length-aware shift mapping is to find an additional mapping $S: \ybar \mapsto \ybar'$ that further perturbs the offset path $\ybar$ into a new path $\ybar'$ satisfying a prescribed path-length condition
\begin{equation}
\label{eq:path_length_constraint}
    \pathLength{\ybar'} = \pathLength{\xbar} + \pathLengthDelta,
\end{equation} where $\pathLengthDelta$ is a constant determined by the reuse configuration (e.g., $\pathLengthDelta = 0$ for reuse within the same time gate, or $\pathLengthDelta = \Delta t$ for time-gate shifting).
Since we perform the transformation starting from $\ybar$, it is convenient to express the constraint as
\begin{equation}
    \pathLength{\ybar'} = \pathLength{\ybar} + \pathLengthDelta,
\end{equation}
where $\pathLengthDelta$ is a fixed constant given $\xbar$; we retain this notation by a slight abuse.
For simplicity, we assume that $\xbar$ and $\ybar$ share the endpoints adjacent to the reconnection vertex, so that $\pathLengthDelta$ is independent of its position; the general case is detailed in the supplementary material.
Rather than modifying the entire offset path $\ybar$, we only adjust the reconnection vertex $\pbf$, following a strategy similar to the localized mutation used by \citet{Sawhney:2022:Decorrelating}.
This further simplifies the constraint into 
\begin{equation}
    \pathLength{\pbf'} = \pathLength{\pbf} + \pathLengthDelta,
\end{equation}
where $\pathLength{\pbf}$ is now path length sum from two adjacent vertices $\pbf_1, \pbf_2$, on the offset path (corresponds to $\ybf_3, \xbf_5$ in \cref{figure:path_length_shiftmap_overview}).
Because our transformation is deterministic, we refer to it as a mapping rather than a mutation.

\subsection{Two-Dimensional Surface Parameterization}
\begin{figure}
\includegraphics[width=0.85\linewidth]{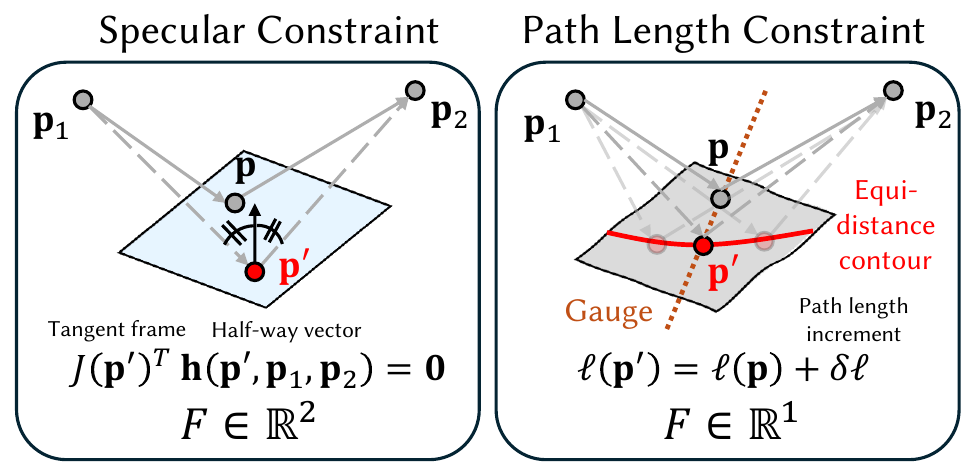}
\caption{Specular material imposes a 2D constraint, which uniquely determines $\pbf'$ (function of $\xi'$) on the surface given $\pbf_1$ and $\pbf_2$. On the other hand, path length imposes a 1D constraint, the feasible $\pbf'$ lies on an equi-distance contour on the surface. To further reduce the degree of freedom, we introduce an additional constraint, referred to as a gauge in this paper.}
\label{figure:specular_path_length_comparison}
\end{figure}

We focus on surface-based light transport and parameterize the reconnection vertex $\pbf$ using a two-dimensional coordinate $\xi \in \mathbb{R}^2$.
This parameterization may take various forms, such as barycentric coordinates on a triangle, uv coordinates of an implicit surface, or polar coordinates on a hemispherical shadow map.
Our objective is to find a new surface coordinate $\xi'$ that satisfies the path-length constraint:
\begin{equation}
    \pathLength{\xi'} = \pathLength{\xi} + \pathLengthDelta.
\end{equation}
Mathematically, this problem closely resembles the computation of valid paths on thin (or delta) manifolds induced by specular reflection~\cite{Jakob:2013:Light, Kaplanyan:2014:Naturalconstraint, Lehtinen:2013:Gradientdomain}.
As in prior work, we therefore employ Newton’s method to solve for $\xi'$.
However, there is a fundamental difference between specular manifolds and the path-length manifold considered in ToF rendering.
Specular constraints restrict two degrees of freedom from halfway vector alignment, which locally determine a unique solution for $\xi'$. 
In contrast, the path-length constraint introduces only a single scalar constraint, leaving one remaining degree of freedom in $\xi'$ (\cref{figure:specular_path_length_comparison}). 
As a result, $\xi'$ is not uniquely defined under the path-length constraint alone.


{To resolve this ambiguity, we introduce an additional constraint, which we refer to as a \emph{gauge condition} by analogy. 
This terminology reflects the similarity to gauge fixing in physics, where redundant degrees of freedom are removed by enforcing invariance under a chosen constraint, although there is no direct physical connection to electromagnetism.
Assuming that a chosen gauge function $\gauge{\cdot}$ remains invariant under the shift, we can define a two-dimensional constraint function $F$ as
\begin{equation}
    F(\xi,\xi') =
    \begin{bmatrix}
        \pathLength{\xi'} - \pathLength{\xi} - \pathLengthDelta \\
        \gauge{\xi'} - \gauge{\xi}
    \end{bmatrix}
    = \begin{bmatrix}0\\0\end{bmatrix}.
\end{equation}
This formulation yields a well-posed system that can be solved efficiently using Newton iteration.

\subsection{Selecting a Good Gauge Condition}
A simple choice of gauge is a fixed linear axis, e.g., $G(\xi) = a^T \xi$ (\cref{figure:gauge_comparison_figure}-(A)).
This choice is reversible and easy to implement.
However, it often results in large displacements in surface parameter space, which can negatively impact path similarity.
To reduce deformation, a natural alternative is to move along the gradient direction
$\nabla \pathLength{\xi}$ which locally minimizes vertex displacement (\cref{figure:gauge_comparison_figure}-(B)).
However, this approach is not reversible: starting from $\xi'$ and reducing the path length by $\pathLengthDelta$ does not recover $\xi$ unless $\nabla \pathLength{\xi} \mathbin{\|} \nabla \pathLength{\xi'}$.
This lack of bijectivity violates the requirements for unbiased shift mapping.


Instead, we propose a bijective average-gradient gauge (\cref{figure:gauge_comparison_figure}-(C)) which symmetrically depends on both the start and end points. 
Specifically, we define the constraint function
\begin{equation}
    F(\xi,\xi') =
    \begin{bmatrix}
        \pathLength{\xi'} - \pathLength{\xi} - \pathLengthDelta \\
       (Rm)^T (\xi' - \xi)
    \end{bmatrix}
    = \begin{bmatrix}0\\0\end{bmatrix},
\end{equation}
where $m = \cP{\nabla \pathLength{\xi} + \nabla \pathLength{\xi'}} / 2$ and $R$ denotes a 90 degree rotation matrix.
Because the search direction $m$ depends equally on $\xi, \xi'$, this formulation is symmetric and therefore reversible, ensuring bijectivity and unbiasedness of shift-mapping.
Practically, in many cases, we could not easily see the biasedness of sample reuse using the initial gradient only, as the moving distance is typically small compared to the gradient variation.
However, to maintain theoretical correctness, we choose the average gradient method throughout the manuscript.
Meanwhile, we can also use mid-point gradient $m = \nabla\pathLength{(\xi + \xi')/2}$, and this behaves similarly to using average gradient.
However, since our objective is to solve for the endpoint $\xi'$, the midpoint method still requires evaluating a path at $\xi'$, which introduces additional overhead.

\begin{figure}
\includegraphics[width=\linewidth]{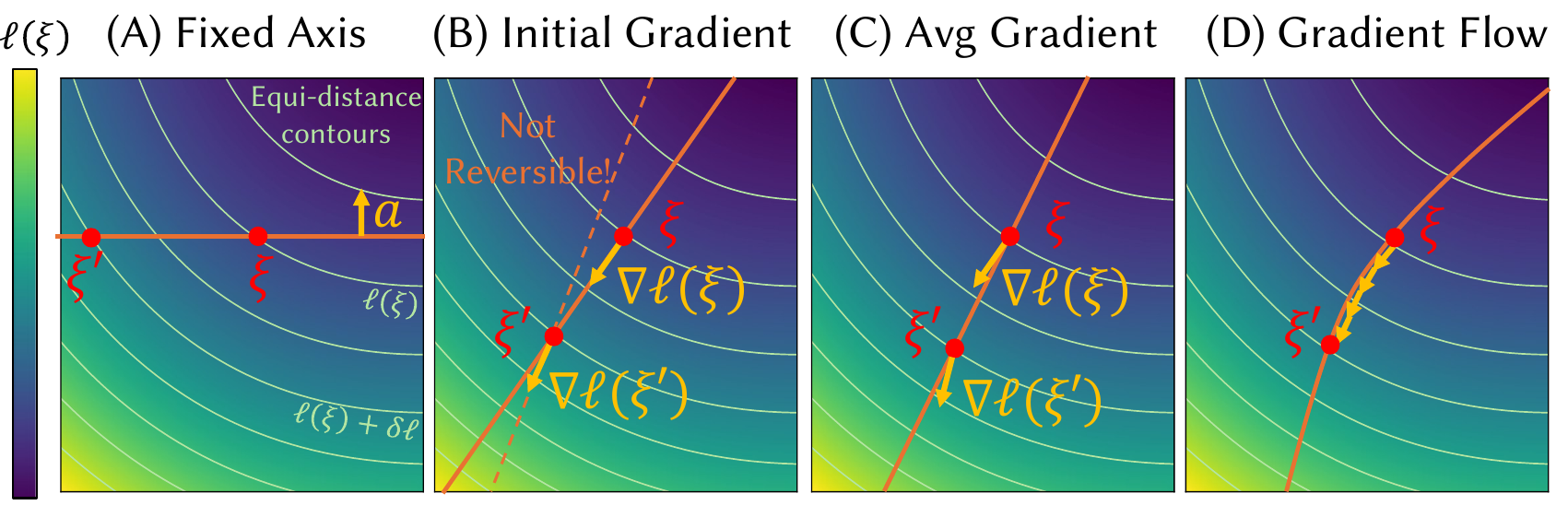}
\caption{
Methods for imposing an additional constraint.
(A) A fixed-axis gauge is reversible and easy to implement,
but it may require traveling a long distance to find a valid solution.
(B) Tracing along the gradient reduces the displacement,
but it is not guaranteed to be reversible.
(C) Our average-gradient method uses gradients at both the start and end points, ensuring reversibility.
(D) Gradient flow is also bijective and yields small displacements, but it is not based on Newton’s method and requires many small steps.
}
\label{figure:gauge_comparison_figure}
\end{figure}

 
\paragraph{Newton's Iteration} Now we solve for $\xi'$ using Newton’s method, initialized at $\xi$ and iteratively updated using gradients:
\begin{equation}
    \xi'_0 = \xi, \hspace{5mm} \xi'_{k+1} \leftarrow \xi'_{k} - \cP{\frac{\partial F}{\partial \xi'_{k}}}^{-1} F(\xi, \xi'_{k}).
\end{equation}
We use backtracking line search as \citet{Jakob:2012:Manifold} for better step-size control.
We also require the determinant of the Jacobian of the shift mapping, which can be computed via implicit differentiation as
\begin{equation}
    \det\cP{\frac{\partial\xi'}{\partial\xi}} = \det \cP{ \frac{\partial F}{\partial\xi}} \Biggm/ \det \cP{ \frac{\partial F}{\partial\xi'}}.
\end{equation}
Note that we still need to consider an additional Jacobian term that transforms $\xi$ to $\pbf$ to be applied to \cref{eq:GRIS_resampling_weight,eq:GRIS_MIS_weight}:
\begin{equation}
    \det\cP{\frac{\partial\pbf'}{\partial\pbf}} = \frac{\sqrt{\det J'^TJ'}}{\sqrt{\det J^TJ}} \det\cP{\frac{\partial\xi'}{\partial\xi}},
\end{equation}
where  $J = (\partial \pbf / \partial \xi) \in \mathbb{R}^{3\times2}$ is the \emph{local tangent frame}.

Both Newton's iteration and Jacobian evaluation require the partial derivatives of
$F$ with respect to $\xi$ and $\xi'$ :
\begin{equation}
\begin{aligned}
    \frac{\partial F}{\partial \xi'} &=
    \begin{bmatrix}
        \nabla \pathLength{\xi'} \\
        \cP{Rm + 0.5\nabla^2\pathLength{\xi'}R(\xi'-\xi)}^T
    \end{bmatrix}, \\
    \frac{\partial F}{\partial \xi} &=
    \begin{bmatrix}
        -\nabla \pathLength{\xi} \\
        -\cP{Rm + 0.5\nabla^2\pathLength{\xi}R(\xi'-\xi)}^T
    \end{bmatrix},
\end{aligned}
\end{equation}
which involve both the gradient and the Hessian of $\pathLengthS$.

\paragraph{Gradient and Hessian Evaluation}
The gradient and Hessian of $\pathLengthS$ can be evaluated by transforming derivatives with respect to $\pbf\in\mathbb{R}^3 $ into derivatives with respect to $\xi\in\mathbb{R}^2$.
The path length is given by $\pathLength{\pbf} = \len{\pbf_1 - \pbf} + \len{\pbf_2 - \pbf}$, and we can evaluate its gradient as
\begin{equation}
    \nabla_{\pbf} \pathLengthS = -\frac{\pbf_1 - \pbf}{\len{\pbf_1 - \pbf}} - \frac{\pbf_2 - \pbf}{\len{\pbf_2 - \pbf}} = -(\dbf_1 + \dbf_2) = -\sbf,
\end{equation}
where $\dbf_1, \dbf_2$ are unit direction vectors and $\sbf$ is the unnormalized half-vector.
The Hessian of $\pathLengthS$ with respect to $\pbf$ is
\begin{equation}
    \nabla^2_{\pbf} \pathLengthS = \frac{I - \dbf_1 \dbf_1^T}{\len{\pbf_1-\pbf}} + \frac{I - \dbf_2 \dbf_2^T}{\len{\pbf_2-\pbf}} = A.
\end{equation}
We can then transform the gradient with respect to $\xi$, using the local tangent frame $J$,
\begin{equation}
    \nabla_{\xi} \pathLengthS = J^T (-\sbf).
\end{equation}
Similarly, the Hessian with respect to $\xi$ is given by
\begin{equation}
    \nabla^2_{\xi} \pathLengthS = J^TAJ + Q,
\end{equation}
where $Q$ is defined as
\begin{equation}
   Q \in \mathbb{R}^{2\times2}, \hspace{5mm} Q_{ij} = (-\sbf)^T \cP{\frac{\partial^2 \pbf}{\partial \xi_i \partial \xi_j}}
\end{equation}
for $i, j \in \{1, 2\}$.
The second-order term $\partial^2 \pbf/\partial \xi_i \partial \xi_j$, which captures surface curvature, can be evaluated for specific surface parameterizations.
In practice, we ignore the physical surface curvature, effectively treating the geometry as a collection of infinitesimal flat patches, which simplifies evaluating $Q$.



\paragraph{Comparison with Gradient Flow}
Another intuitive alternative for handling the additional degrees of freedom is to follow the continuous \emph{gradient flow} $\xi(s)$, defined by $\mathrm{d} \xi / \mathrm{d} s = \nabla L (\xi(s))$ (\cref{figure:gauge_comparison_figure}-(D)).
This approach also guarantees bijectivity, but it is significantly slower than Newton’s method (linear vs.\ quadratic convergence).
Further discussion can be found in the supplementary material.

\paragraph{Cost of Newton's Iteration}
In prior steady-state specular manifold methods, the cost of Newton’s iteration is a major concern, as it often requires tracing long specular chains via ray tracing.
In contrast, for the path-length constraint in our setting, it suffices to perturb a single path vertex.
This significantly reduces the cost of each Newton iteration, making it much less expensive than the specular-manifold case.
We note that this strategy does not directly extend to mixed manifolds involving both specular and path-length constraints, which we do not consider in this work.

\subsection{Initial Sample Generation}
So far, we have focused on the efficient reuse of valid samples using path-length-aware shift mapping.
However, time-gated rendering also suffers from a low probability of generating valid initial samples, as the time gate accepts only a very narrow range of path lengths.
In this subsection, we discuss strategies to address this issue.

\subsubsection{Initialization with Ellipsoidal Path Connection}
One approach is to adopt existing techniques that explicitly target valid sample generation for time-gated rendering,
such as ellipsoidal path connection~\cite{Pediredla:2019:Ellipsoidal}.
Our path-length-aware shift mapping is complementary to these techniques,
and can therefore be combined with them to improve the quality of initial RIS samples.
Some adjustments are required to make this approach suitable for GPU implementation,
which we discuss in the implementation section.
\subsubsection{Initial Sampling with Path-length Shrink Mapping}
In some cases, ellipsoidal path connection is not suitable, either because it is too slow or because the geometry is not represented as a triangular mesh.
To address this, we additionally use path-length-aware shift mapping for initial sample generation.
The key idea is simple: we first sample paths using a wider time gate, and then shift (or \emph{shrink}) these samples into a narrower time gate.
In this setting, the path-length constraint in \cref{eq:path_length_constraint} becomes
\begin{equation}
    \pathLength{\xbar'} = \frac{1}{K}\cP{\pathLength{\xbar} - \pathLengthS_0} + \pathLengthS_0,
\end{equation}
where we use $\xbar'$ instead of $\ybar'$ since no baseline shift mapping is required.
Here, $\pathLengthS_0$ is the path length corresponding to the center of the time gate, and $K$ denotes the scaling factor of the wider time gate.
The difference between translation-based mapping (used for spatio-temporal reuse) and shrink mapping (used for initial sample refinement) is illustrated in \cref{figure:fine_tune_shrink_map}.
We can therefore reuse the previous Newton iteration by setting $\pathLengthDelta = \frac{1}{K}\cP{\pathLength{\xbar} -\pathLengthS_0} + \pathLengthS_0 - \pathLength{\xbar}$.
The only required modification is to scale the Jacobian determinant by $1/K$.
Importantly, note that we do not apply this shrink mapping for all RIS samples, but only to the RIS winner, so the computational cost is minimal.

Initial sample generation with shrink mapping can introduce bias if used in isolation. This arises for two main reasons: (1) Newton’s method may become unstable when initial samples are far from the target manifold, and (2) the support of the transformed paths may be smaller than the valid time gate, causing the Jacobian correction (scaling by $1/K$) to be over-applied, which typically leads to darker estimates.
This bias can be eliminated by combining shrink-mapped samples with canonical samples from the target (narrow) time gate using proper MIS weighting (\cref{eq:GRIS_MIS_weight}), which yields an unbiased estimator. In practice, however, we find that allocating all samples to the wider time gate often provides stronger variance reduction, which may outweigh the bias introduced by omitting canonical samples.
Further discussion is provided in the results section.

\begin{figure}
\includegraphics[width=0.95\linewidth]{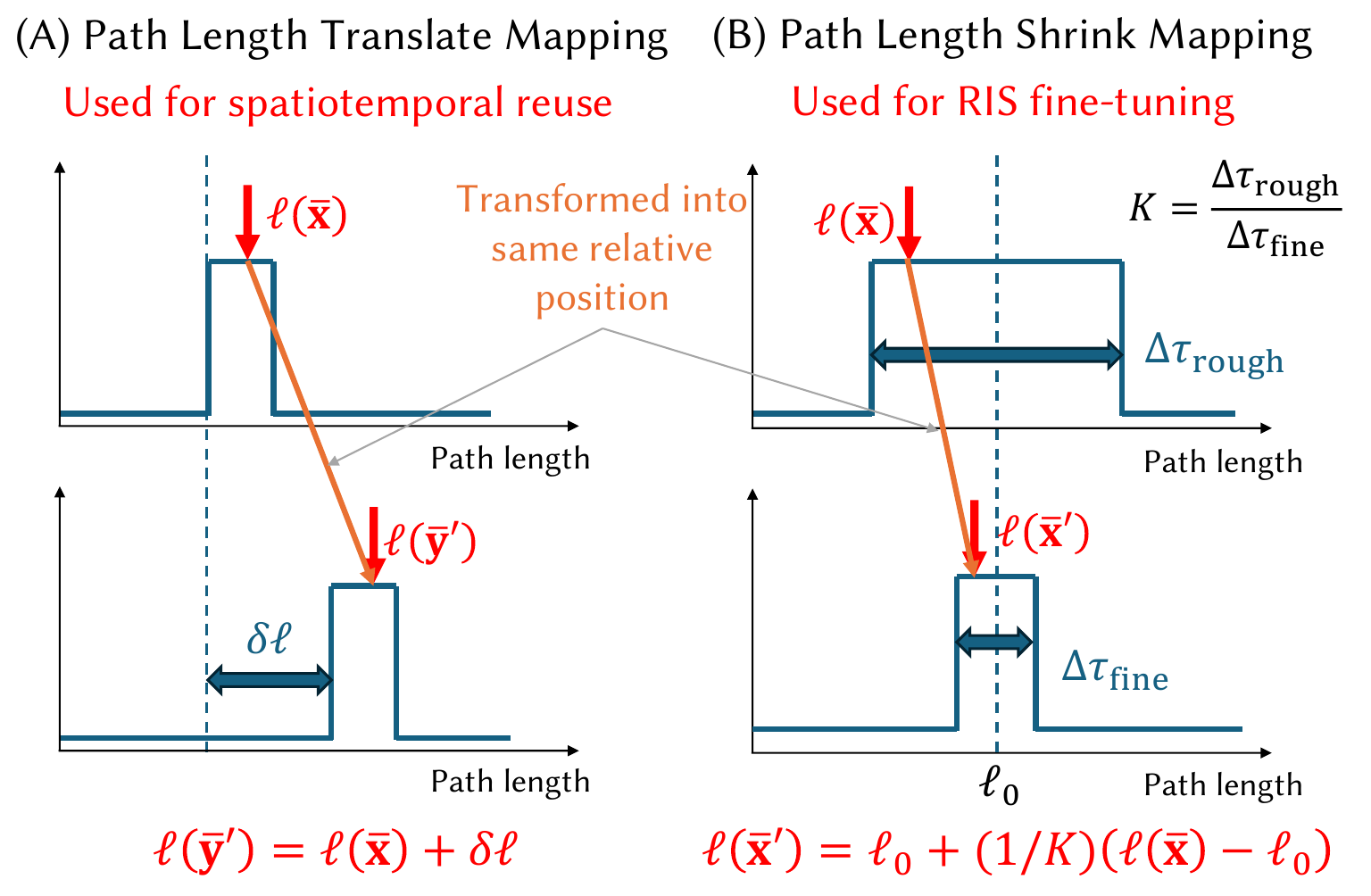}
\caption{
Two types of path-length-aware shift mappings are used for different purposes.
(A) Path-length translation mapping is used for spatio-temporal path reuse.
(B) Path-length shrink mapping is used for efficient initial sample generation: samples are drawn from a wider time gate and then shrunk back to the original gate.
}
\label{figure:fine_tune_shrink_map}
\end{figure}



\subsection{Extension to Transient Histogram Rendering}
\label{sec:transient_histogram_rendering}
Our method is applicable not only to time-gated image rendering, which produces a two-dimensional image ($H\times W$), but also to transient histogram rendering, which outputs a three-dimensional transient histogram ($H\times W \times B$) per frame.
By interpreting transient histogram as a sequence of time gates, we can directly apply path-length-aware shift mapping independently to each histogram bin.
Compared to the image case, histogram rendering introduces two differences: (1) all $B$ reservoirs must be updated for each pixel, increasing the reuse overhead, and (2) there is an additional reuse dimension along the bin axis $B$.
An overview comparing path reuse for image and histogram rendering is shown in~\cref{figure:transient_histogram_reuse}.


\paragraph{Cost-to-Performance Analysis}
As mentioned earlier, a key challenge of path reuse in the histogram setting is the cost-to-performance trade-off, as reuse can be expensive.
Here, we provide a simple analysis to illustrate this effect.
Let the cost of tracing a path sample be $C_\text{trace}$, and the cost of performing a path reuse operation be $C_\text{reuse}$.
For image rendering with path reuse, the total cost is  $C_\text{trace} + C_\text{reuse}$.
For transient histogram rendering, reuse must be applied independently to each bin, increasing the reuse cost by a factor of $B$, while the per-bin quality improvement remains unchanged.
As a result, the relative cost-to-performance efficiency degrades by a factor of
\begin{equation}
    \frac{C_\text{trace} + C_\text{reuse}}{C_\text{trace} + B \times C_\text{reuse}}.
\end{equation}
As $B$ increases, the effectiveness of path reuse therefore decreases.
This effect is particularly pronounced when $C_\text{reuse}$ is large, for example, when path reuse requires expensive path throughput recomputation.
Refer to the results section for a detailed comparison.

\begin{figure}
\includegraphics[width=\linewidth]{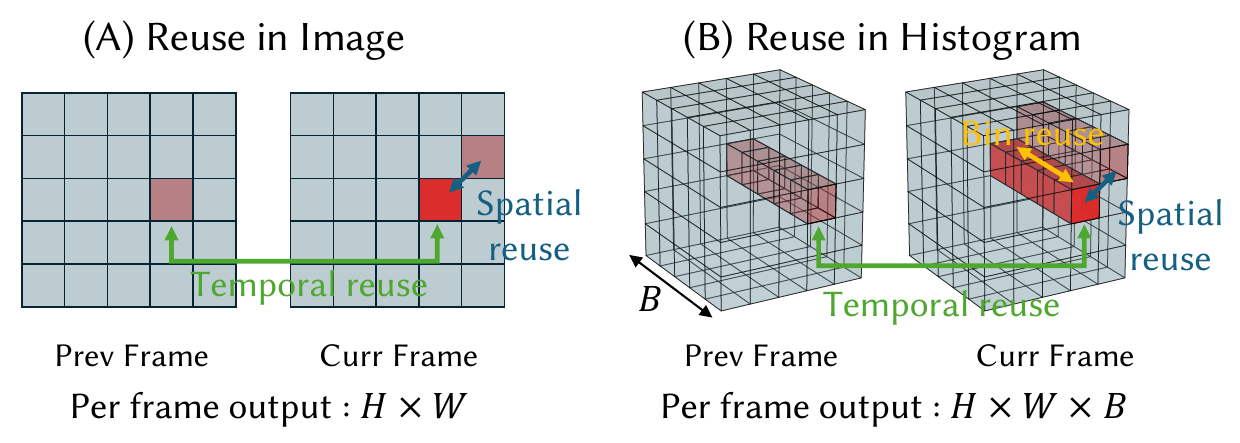}
\caption{
(A) Time-gated image rendering whose per-frame output is $H\times W$, we can use temporal and spatial reuse.
(B) Transient histogram rendering whose per-frame output is $H\times W \times B$, has two differences: (1) reservoir update is more expensive as we need to update all $B$ reservoirs per pixel, and (2) we can use not only temporal and spatial reuse but also bin reuse. 
}
\label{figure:transient_histogram_reuse}
\end{figure}

\paragraph{Comparison with \citet{Jarabo:2014:Framework}'s `Path Reuse'}
For initial sample generation, we adopt the “path reuse” in ~\citet{Jarabo:2014:Framework} (which differs from ReSTIR-style path reuse), even though our overall framework uses ReSTIR.
Specifically, we deposit each sample’s contribution into its corresponding time bin and use these as initial samples for subsequent ReSTIR-style reuse.
We consider the transient histogram as a sequence of time-gated images and apply reuse independently for each bin \emph{only} during the ReSTIR-style path reuse.



%


\subsection{Extension to Doppler Frequency Rendering}
The path-length constraints introduced in the previous sections can be naturally extended to the frequency domain induced by the Doppler effect.
Analogous to the time-resolved path integral in~\cref{eq:tof_path_integral}, the frequency-resolved power spectral density (PSD) $S(\nu)$ resulting from Doppler shift~\cite{kim2025monte} is given by
\begin{equation}
    S(\nu) = \int_\pathspace W_{\nu}\cP{\omega(\xbar)} f(\xbar) \dxbar.
\end{equation}
Here $W_{\nu}$ is a frequency weight similar to $W_\tau$, and $\omega(\xbar)$ denotes the path frequency derived from the \emph{path velocity} $u(\xbar) = -\ddt \pathLength{\xbar}$. This can be expressed as
\begin{equation}
\label{eq:optical_path_velocity}
    u(\xbar) = \sum_{k=0}^{K-1} \eta_{k}\cP{\mathbf{v}_{k} - \mathbf{v}_{k+1}}\cdot \rayvec_{k},
\end{equation}
where $\rayvec_{k}$ is the unit vector from $\xbf_k$ to $\xbf_{k+1}$ and $\mathbf{v}_{k} = \ddt \mathbf{x}_{k}$ denotes the instantaneous velocity at each vertex.
The path velocity induces a Doppler frequency shift given by $\Delta f(\xbar)\coloneq f_0 u(\xbar) / c$, where $f_0$ is the carrier frequency\footnote{For chirped lasers, $\omega$ depends on both path length and velocity. Here, we consider only a standard (constant-frequency) laser, for which $\omega(\xbar) = 2\pi \Delta f(\xbar)$.}.


Rather than constraining the optical path length $\pathLength{\pbf}$, we impose a constraint on its temporal derivative $\ddt \pathLength{\pbf}$ i.e., path velocity.
This constraint defines a different manifold from the path-length manifold considered earlier.
We visualize examples of manifolds induced by path-velocity constraints in~\cref{figure:velocity_manifold} for three different cases.
A comparison between time-gated and frequency-gated rendering is also summarized in~\cref{table:time_vs_frequency_gate}.
Overall, the same path reuse framework can be applied by substituting the path-length constraint with the path-velocity constraint.
Full derivations of the corresponding gradients and Hessians are provided in the supplementary material.


\begin{figure}
\includegraphics[width=\linewidth]{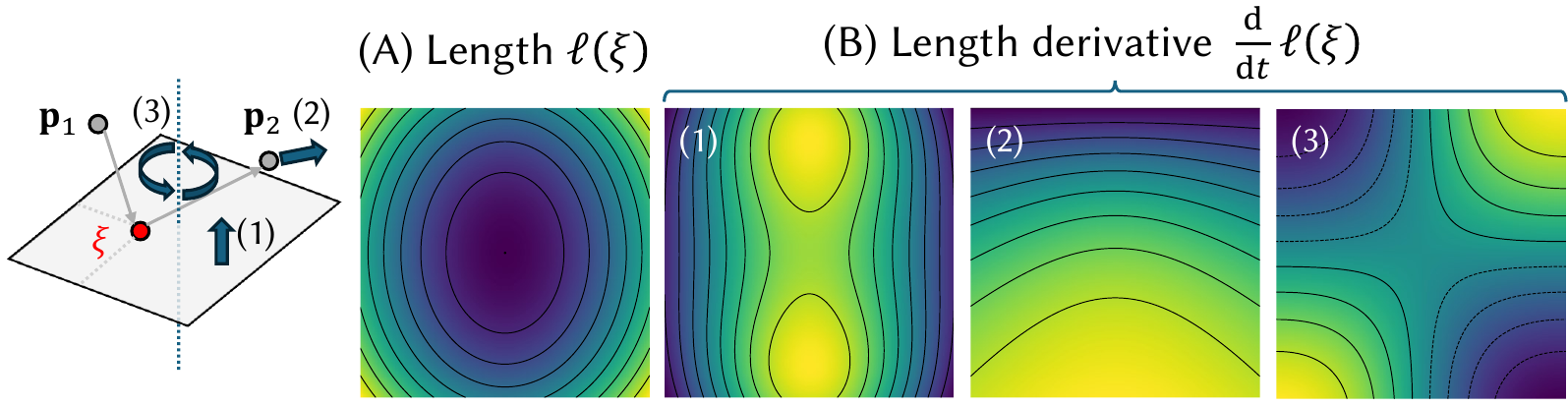}
\caption{
(A) Path length $\pathLength{\xi}$ on the plane, forming ellipsoidal iso-value contours.
(B) Temporal derivative of path length, $\ddt \pathLength{\xi}$, on the plane under three different scenarios: (1) the plane moves upward, (2) $\pbf_2$ moves away, and (3) the plane rotates, each producing distinct iso-value contours.
}
\label{figure:velocity_manifold}
\end{figure}


\begin{table}[h]
\centering
\caption{Comparison between time-domain and frequency-domain formulations and their corresponding constraints.}
\label{table:time_vs_frequency_gate}
\small
\begin{tabular}{lcc}
\toprule
 & \textbf{Time Domain} & \textbf{Frequency Domain} \\
\midrule
Measurement & Transient signal $I(\tau)$ & PSD $S(\nu)$ \\
Depends on & Path Length & Path Velocity \\
Notation & $\pathLength{\xbar}$ & $u(\xbar)= - \ddt \pathLength{\xbar}$ \\
Constraint & $\pathLength{\pbf'} = \pathLength{\pbf'} + \pathLengthDelta$ & $\ddt \pathLength{\pbf'} = \ddt \pathLength{\pbf'} + \delta u$ \\
\bottomrule
\end{tabular}
\end{table}

%% file: sections/05_implementation_v1.1.tex
\section{Implementation Detail}
\begin{figure}
\includegraphics[width=0.9\linewidth]{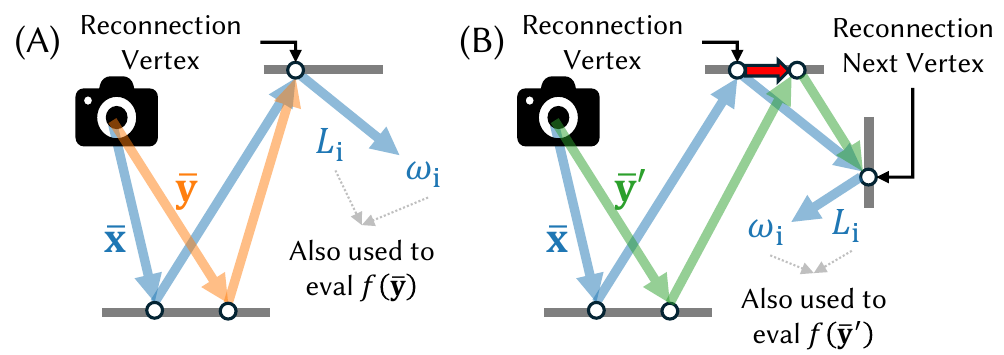}
\caption{
(A) \citet{Lin:2022:Generalized} stores the hit record of the reconnection vertex along with additional information (incident direction $\omega_\mathrm{i}$ and radiance $L_\mathrm{i}$) to recompute the offset path throughput.
(B) We additionally store the next vertex after the reconnection vertex and attach this information to that vertex instead of the original reconnection vertex.
The red arrow indicates path-length-aware shift mapping.
}
\label{figure:reconnection_vertex_impl}
\end{figure}

\paragraph{Reservoir Implementation}
Our reservoir structure largely follows the design of \citet{Lin:2022:Generalized}.
The main difference is that we additionally store the next vertex of the reconnection vertex, and keep the reconnection information (e.g., incident direction and radiance used to recompute the path throughput) at this next vertex rather than at the original reconnection vertex.
This difference is illustrated in \cref{figure:reconnection_vertex_impl}.
Note that for temporal reuse under dynamic lighting or animated meshes, the reconnection information must be re-evaluated to remain valid.
For further details, see \citet{Lin:2022:Generalized}.




\paragraph{Light Source}
Rather than considering general scene lighting, we restrict our setup to delta light sources, which are commonly used in ToF applications. Specifically, we model a spotlight under two regimes: (1) a collimated, laser-like case with a very small angular extent, and (2) a point light–like case with a large angular extent.
In the collimated case, we trace additional rays from the light source to generate light samples prior to performing next event estimation (NEE), similar to bidirectional path tracing.
In the point light case, we instead rely on direct NEE.

\paragraph{Ellipsoidal Path Connection}
Several modifications are required to adapt the original ellipsoidal path connection method~\cite{Pediredla:2019:Ellipsoidal} to our setting. In the original approach, ellipsoid–AABB and ellipsoid–triangle intersection tests are performed on the BVH to gather all triangles that may intersect a given ellipsoid. However, this strategy does not scale well on GPUs. For complex meshes, it can produce thousands of candidate triangles, leading to excessive memory usage and register spilling.
Instead of explicitly collecting all intersecting triangles, we sample BVH nodes using heuristic weights. Each node is assigned a weight proportional to the total area of the triangles it contains, and set to zero weight if it does not intersect the ellipsoid. We also apply MIS between direct connection and ellipsoidal path connection to robustly handle specular materials.
Finally, to avoid the complexity of a full BDPT implementation, we restrict ellipsoidal path connection to NEE only.

%% file: sections/06_results_v1.2.tex
\section{Results}
We implement our algorithm within the Falcor framework~\cite{Kallweit22} and conduct experiments with an AMD Ryzen Pro 3955WX CPU and an NVIDIA RTX 3090 GPU.
Most test scenes are taken from~\citet{Bitterli:2016:Rendering}; however, to better mimic laser-based ToF sensing, we replace the original scene light sources with a custom delta light source.
We report and visualize only indirect illumination, as direct illumination under a delta light source is trivial.

\subsection{Offline Time-Gated Rendering with Spatial Reuse}
We first present offline time-gated rendering results for the \textsc{Cornell-Box} scene with different material configurations for initial validation.
A collimated laser illuminates the upper region of the back wall, with its direction parallel to the camera direction.
The box size is $2\times2\times2$, and the time gate, which is a box function, is set to cover $0.5-2.5 \%$ of the scene scale.
For the ReSTIR configuration, following \citet{Lin:2022:Generalized}, we use 32 path trees as initial candidates and perform spatial reuse three times, each time considering five neighboring pixels within a 10-pixel radius.
We compare three methods: path tracing without any reuse, naive reuse based on ReSTIR-PT, and our proposed method.
All methods are evaluated under the same time budget.
We report the mean absolute percentage error (MAPE) as the evaluation metric, following \citet{Lin:2022:Generalized}.

\begin{figure} 
\includegraphics[width=0.95\linewidth]{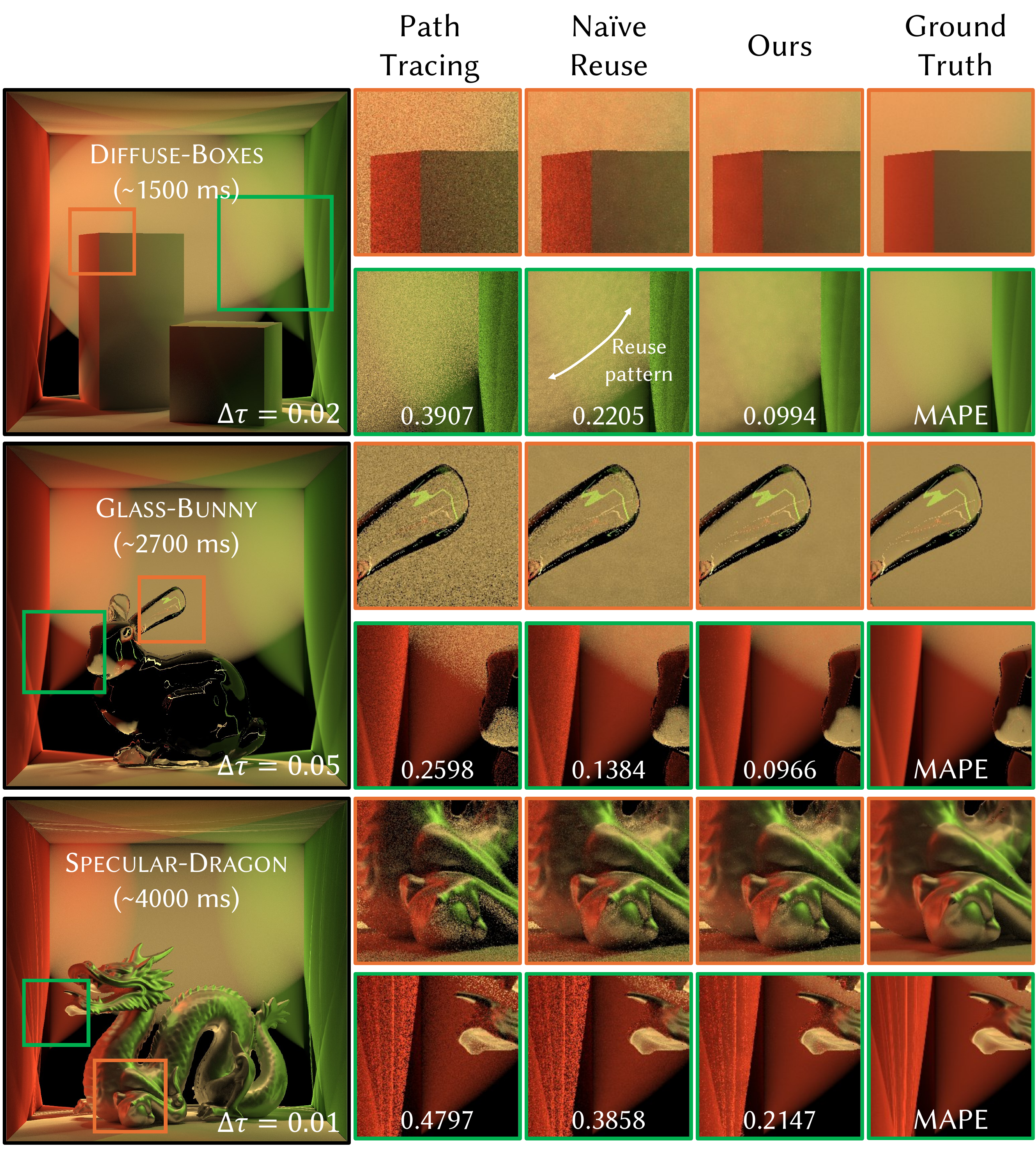}
\caption{Comparison of the \textsc{Cornell-Box} scene with various objects and materials under an equal time budget.
All images are rendered at a resolution of $1024 \times 1024$, with the maximum path depth set to 6.}
\label{figure:cornell_box_diff_materials}
\end{figure}

\begin{figure*} 
\includegraphics[width=1.0\linewidth]{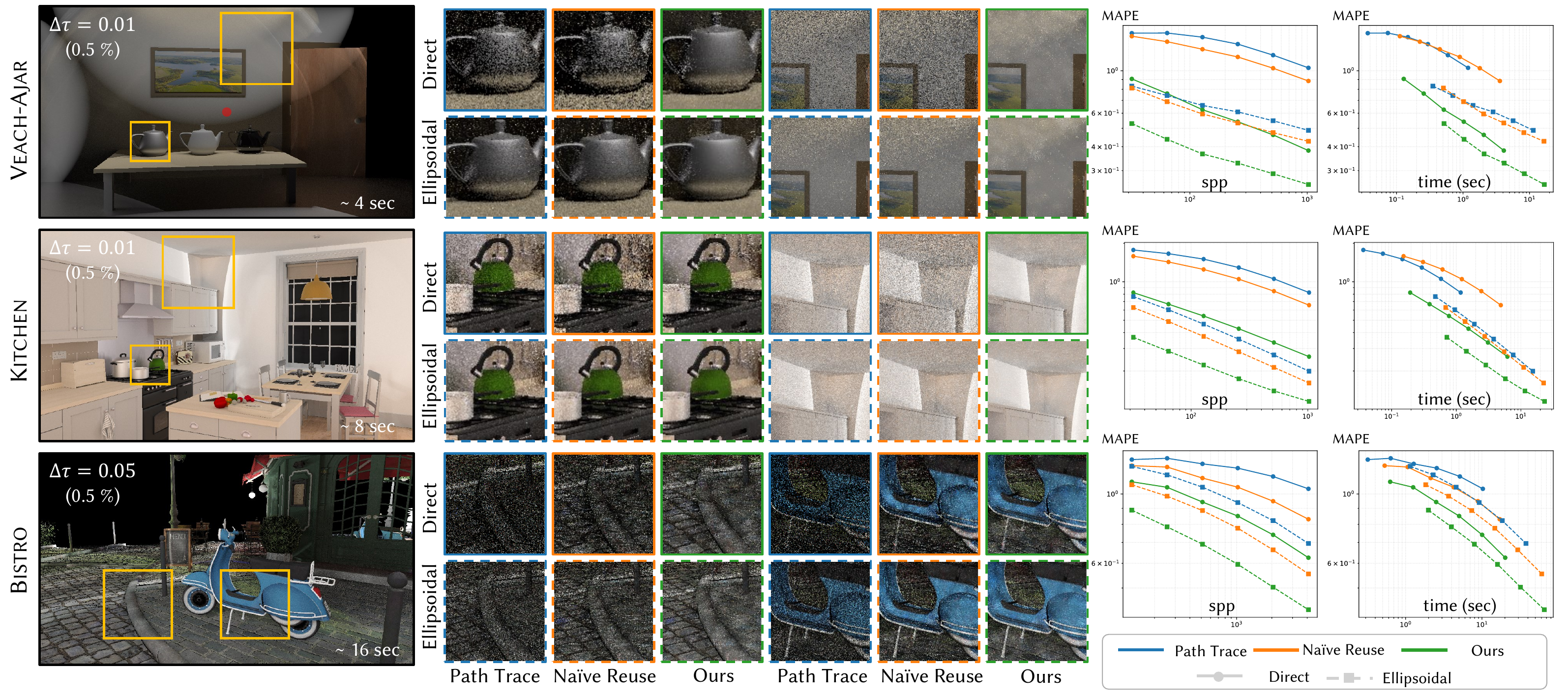}
\caption{We implement ellipsoidal path connection~\cite{Pediredla:2019:Ellipsoidal}, adapted to a GPU environment, and use it as an initial candidate generation.
Compared to a direct connection, an ellipsoidal path connection produces better initial samples at the cost of increased computation time.
Our method improves the result regardless of the initial candidate generation method.
The light source is collocated with the camera.
For the \textsc {Veach-Ajar} scene, a collimated laser is used, while a point light is used for the others. 
Image size is $960 \times 540$, and max bounce is set to 6.
}
\label{figure:ellipsoid_comparion}
\end{figure*}

The results are shown in~\cref{figure:cornell_box_diff_materials}.
Path tracing produces noisy images, while naive reuse moderately improves image quality but still fails to produce satisfactory results.
Interestingly, naive reuse exhibits ellipsoidal reuse patterns (highlighted by the white curved arrow), as valid reuse without explicit path-length preservation can only occur along ellipsoidal loci that satisfy the path-length constraint.
In contrast, our proposed path-length-aware shift mapping produces the best results across different scenarios, as it ensures that reused paths remain valid across neighboring pixels, even when they lie on different ellipsoids.
Results for more scenes, including convergence plots, are shown in~\cref{figure:ellipsoid_comparion}. 
Our method consistently achieves the best performance, while naive reuse can even perform worse than plain path tracing due to its low reuse validity rate.

\begin{figure} 
\includegraphics[width=\linewidth]{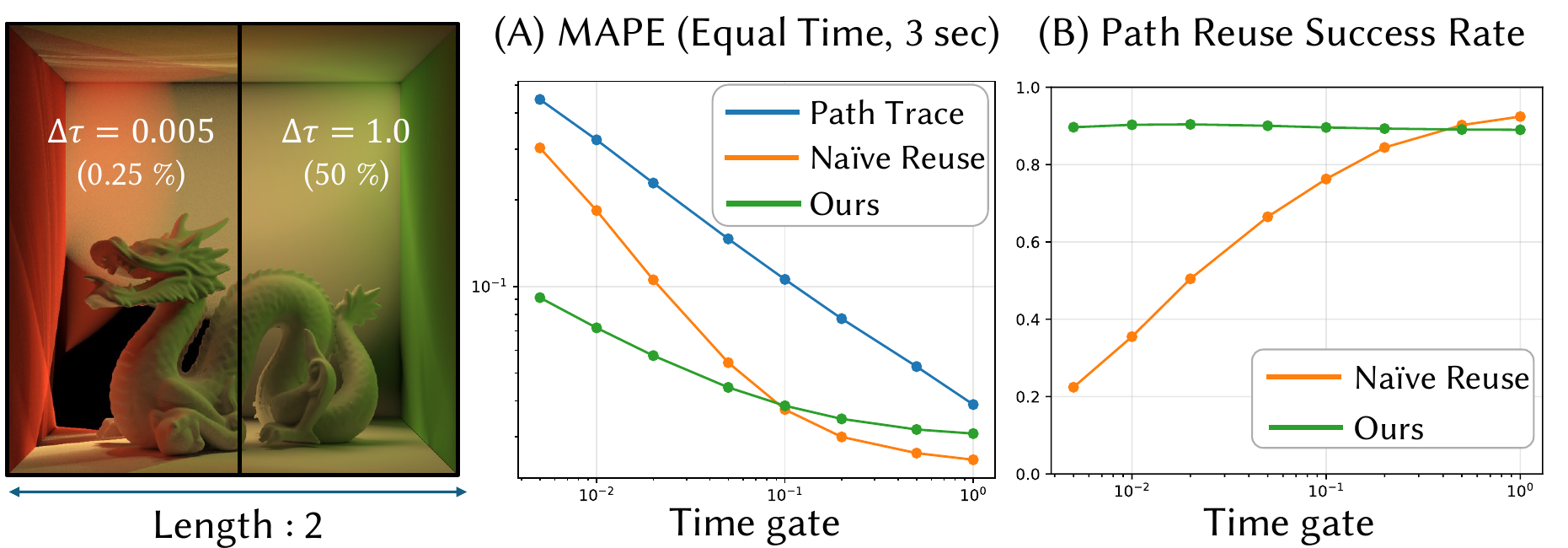}
\caption{(A) MAPE comparison under different time-gate widths within a 3-second time budget.
(B) Comparison of path reuse success rates between naive reuse and our method.
Overall, our method performs best when the time gate is sufficiently narrow, $5-10\%$ of the scene size. }
\label{figure:time_gate_comparison}
\end{figure}

\paragraph{Effect of Time Gate Width}
Similar to~\citet{Pediredla:2019:Ellipsoidal}, we compare results under different time-gate widths $\Delta \tau$ in \cref{figure:time_gate_comparison}.
All images are normalized by the time-gate width to ensure comparable intensity levels.
The error for different $\Delta \tau \in [0.005, 1.0]$ is plotted in \cref{figure:time_gate_comparison}–(A).
We find that our method is most effective for narrow time gates.
As the time gate widens, naive path reuse is less likely to violate the path-length constraint, making the additional computation for path-length correction less beneficial.
In \cref{figure:time_gate_comparison}–(B), we also report the path reuse success rate for both naive reuse and our method.
For very small time gates, naive reuse achieves a very low success rate, which gradually increases as the time gate widens.
In contrast, our method maintains a nearly constant and high reuse success rate across different time-gate widths.
Notably, for very large time gate, the success rate of our method becomes slightly lower than that of naive reuse.
This reduction is primarily due to failures in Newton’s iteration or additional occlusion introduced when moving the reconnection vertex.
Overall, our method is most beneficial for narrow time gates, whereas naive reuse is sufficient for wider gates.
The exact threshold depends on the scene, but we observe that$\Delta \tau$ below approximately $5-10 \%$ of the scene scale serves as a practical boundary.

\begin{figure}
\includegraphics[width=0.90\linewidth]{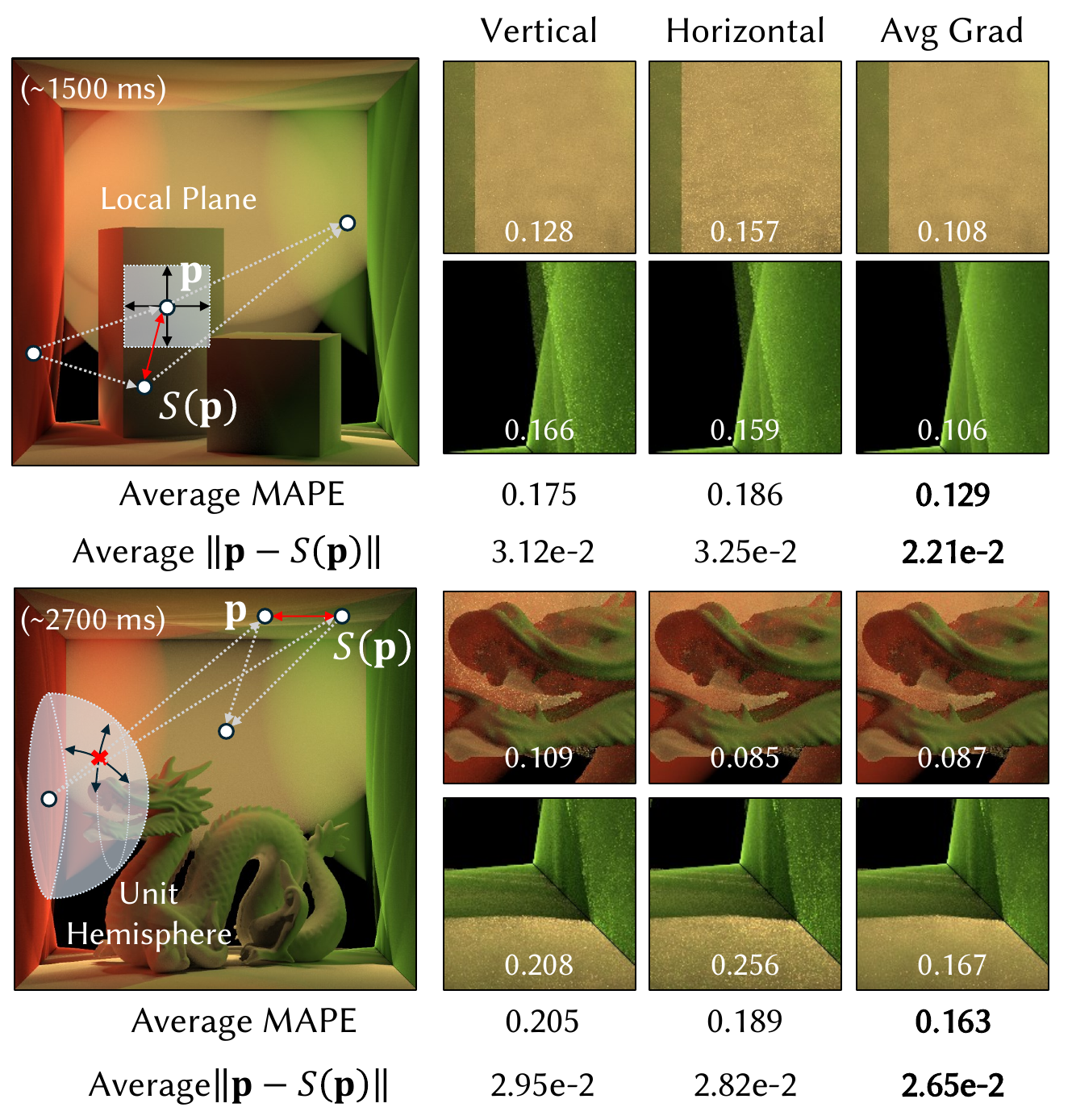}
\caption{Comparison of gauge directions using fixed horizontal and vertical axes, and the proposed average-gradient direction.
Our method consistently produces the best results, largely due to the reduced displacement.}
\label{figure:gauge_axis_comparison}
\end{figure}

\paragraph{Gauge Direction Comparison}
We demonstrate the effectiveness of the proposed gauge in~\cref{figure:gauge_axis_comparison}. We evaluate horizontal, vertical, and average-gradient directions under different coordinate systems, including a local plane and a unit hemisphere\footnote{Each coordinate system has its own advantages and limitations, and we employ heuristics that leverage both. See the supplementary material for details.}. Each method is evaluated under an equal-time budget with $\Delta\tau = 0.01$.
We observe that, for each region, one fixed axis typically performs better than the other in terms of error, depending on the scene geometry. However, regardless of the coordinate system, the gradient-based direction achieves the best performance on average.
We also report the moving distance of the reconnection vertex ($\len{\pbf - S(\pbf)}$) induced by each method and find that our gradient-based approach consistently results in the shortest displacement.
In fact, moving distance is not a perfect indicator of path-throughput similarity, since shorter displacements may break the similarity due to newly appeared occlusions.
Still, we could observe a clear overall correlation between shorter moving distances and reduced variance.
Regarding the performance, we could not find a noticeable difference when evaluating additional Hessians from using the gradient gauge direction.

\paragraph{Newton's Iteration Statistics}
We also report statistics for Newton’s iteration.
We set the convergence threshold to $1\%$ of the time-gate width and limit the maximum number of iterations to 5.
Consistent with the observations of \citet{Jakob:2012:Manifold}, we find that approximately two to three Newton iterations are sufficient to reach a valid solution, with a failure rate below $10-15\%$.
The overall statistics are summarized in~\cref{tab:newton_statistics}.
Here, \emph{Newton SR} denotes the Newton's iteration success rate and \emph{Actual SR} denotes the actual path reuse success rate, which is consistently lower, as it excludes samples that are occluded (from $\pbf_1$) or rejected due to excessive Jacobian factors.
 

\begin{table}[t]
\centering
\caption{Newton iteration counts and success rates across different scenes.
The actual success rate accounts for occlusion and Jacobian factors and is always lower than the Newton success rate.}
\begin{tabular}{lccc}
\hline
Scene & \# of Iter. & Newton SR & Actual SR \\
\hline
\textsc{Cornell-Box}     & 1.5267 & 0.9652 & 0.9356 \\
\textsc{Cornell-Dragon} & 1.6499 & 0.9454 & 0.9027 \\
\textsc{Veach-Ajar}     & 1.5549 & 0.9578 & 0.9026 \\
\textsc{Kitchen}        & 1.9003 & 0.9197 & 0.8127 \\
\textsc{Bedroom}        & 1.9854 & 0.9199 & 0.8331 \\
\textsc{Staircase}      & 2.2179 & 0.8664 & 0.7238 \\
\textsc{Bistro}         & 2.3480 & 0.8455 & 0.6638 \\
\hline
\end{tabular}
\label{tab:newton_statistics}
\end{table}


\paragraph{Initialization with Ellipsoidal Path Connection}
So far, we have used direct connection for initial sample generation, but ellipsoidal path connection~\cite{Pediredla:2019:Ellipsoidal} can also be employed for this purpose, producing higher-quality initial samples.
The results are shown in \cref{figure:ellipsoid_comparion}.
We present equal-time comparisons for different strategies, along with convergence plots with respect to both spp and time.
Compared to direct connection, ellipsoidal connection yields better results, although the improvement depends on the scene.
Importantly, we find that regardless of the initial sampling strategy, our method consistently outperforms the baseline methods.
The computational cost of ellipsoidal path connection is significantly higher than that of direct connection---by a factor of $5\times$ to $10\times$, depending on the scene---which often necessitates the use of narrower time gates to clearly observe its advantage.
Our heuristic-based ellipsoidal sampling strategy may be suboptimal, as it is not the primary focus of this paper.
Designing more efficient ellipsoidal connection algorithms that are well-suited for GPU execution is an interesting direction for future work.

\begin{figure} 
\includegraphics[width=0.95\linewidth]{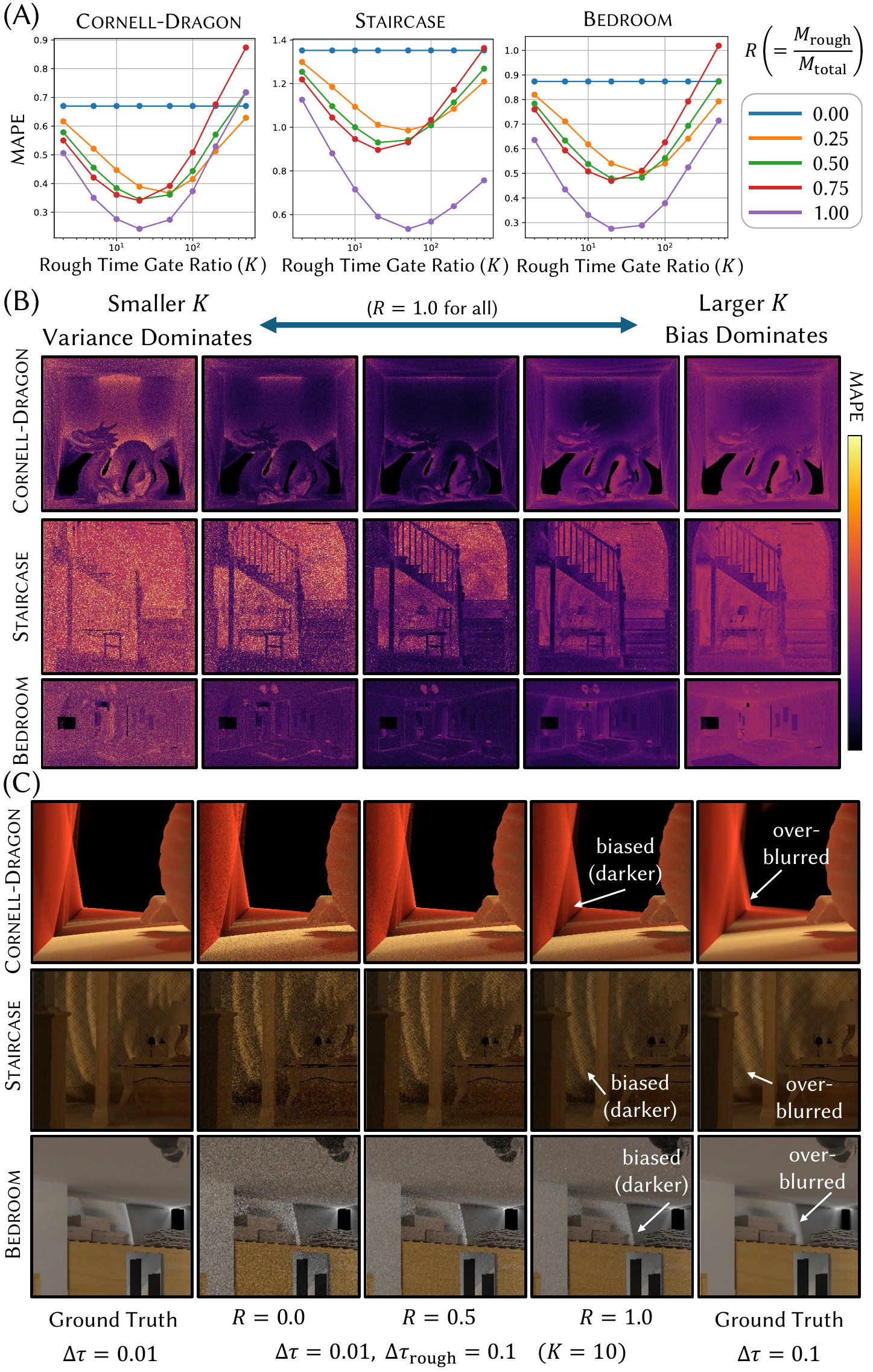}
\caption{
Results of path-length shrink mapping using a rough time gate.
(A) When the rough time gate is too small, little benefit is gained from widening the gate, and variance dominates, whereas excessively large rough time gates introduce bias.
(B) Visually acceptable results are typically obtained when the rough time gate is approximately 10--20 times wider than the fine time gate.
(C) Simply increasing the time-gate width leads to over-blurring of scene details.
}
\label{figure:fine_tune}
\end{figure}

\subsection{Evaluation of Path Length Shrink Mapping}
In this subsection, we evaluate the proposed path-length shrink mapping as an efficient strategy for initial sample generation under narrow time gates.
To clearly show the variance–bias behavior, we disable spatial reuse in this experiment.
We focus on two parameters: the relative width of the rough time gate, defined as $K=\frac{\Delta\tau_{\text{rough}}}{\Delta\tau_{\text{fine}}}$, and the fraction of samples allocated to the rough time gate, defined as $R=\frac{M_{\text{rough}}}{M_{\text{rough}}+M_{\text{fine}}}$.
\cref{figure:fine_tune}–(A) reports the MAPE for different values of $K \in [1.0, 500]$ and $R \in [0.0, 1.0]$, evaluated at a fixed fine time gate of $\Delta\tau = 0.01$.
Overall, we observe that using a moderately larger rough time gate and allocating all samples to the rough gate yields the best metric across different scenes.
In \cref{figure:fine_tune}–(B), we fix $R=1.0$ and vary the value of $K$.
When the rough time gate is too small, the benefit of using a wider gate is limited, and variance dominates the result.
On the other hand, when the rough time gate becomes excessively large, the result becomes biased and typically darker than the ground truth image, since the region mapped from the rough time gate becomes smaller than the original fine gate.
The MAPE plot in \cref{figure:fine_tune}–(A) suggests that an acceptable bias–variance trade-off occurs for $K$ values between approximately $10$ and $100$.
This range is scene dependent and also varies with the chosen time-gate width.

We further examine the effect of sample allocation in \cref{figure:fine_tune}–(C) by varying $R$ while keeping $K$ fixed.
As discussed earlier, setting $R<1$ and applying MIS between canonical samples from the fine time gate and candidate samples from the rough time gate produces an unbiased estimator.
Setting $R=1$ completely ignores canonical samples, resulting in a biased estimate that is typically darker than the reference image, but with substantially reduced variance.
Despite being biased, this approach does not introduce the blurring artifacts observed when simply increasing the time-gate width alone (rightmost column of \cref{figure:fine_tune}–(C)).
This improvement arises because shrink mapping explicitly attempts to match path lengths, whereas merely widening the time gate does not.

\begin{figure} 
\includegraphics[width=\linewidth]{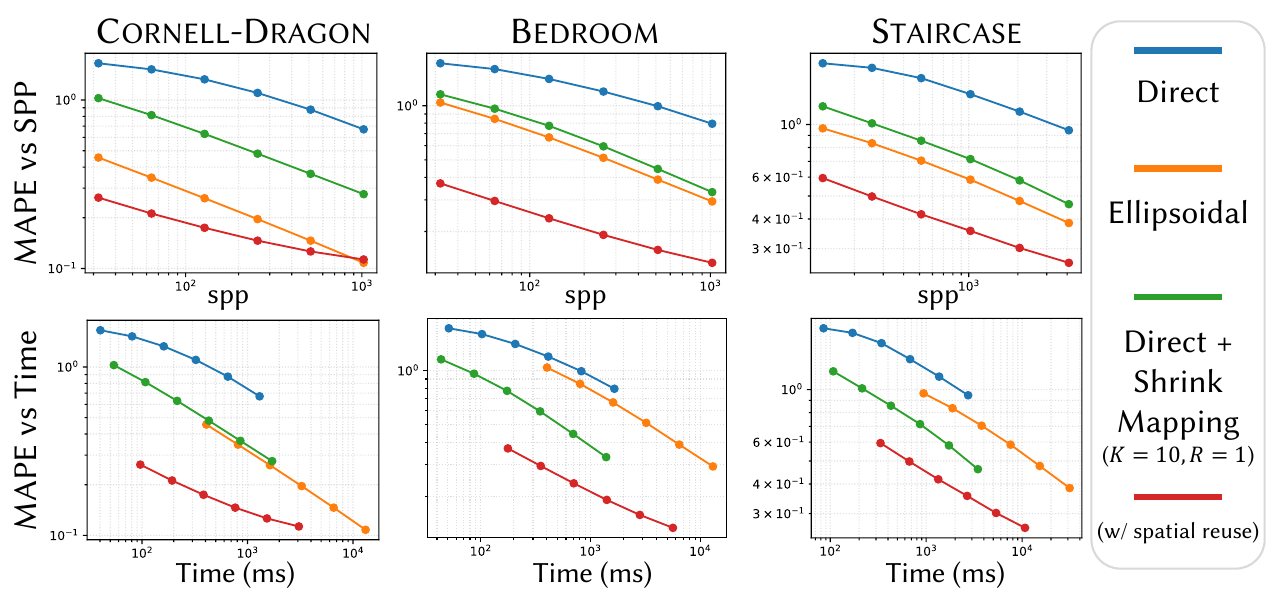}
\caption{Comparison between ellipsoidal path connection, direct connection, and direct connection combined with shrink mapping.
Under an equal-spp constraint, an ellipsoidal path connection achieves the best quality.
However, for complex scenes where ellipsoidal sampling becomes expensive, our method provides a more favorable trade-off under an equal-time budget.
We also show shrink mapping with spatial reuse as a red line.
}
\label{figure:fine_tune_ellipsoid}
\end{figure}

\paragraph{Performance Comparison with Ellipsoidal Path Connection}
Since shrink mapping is designed to address sampling difficulties under narrow time gates without expensive scene–ellipsoid intersections, we compare it with ellipsoidal path connection in \cref{figure:fine_tune_ellipsoid}.
We use $R=1$ and $K=10$.
Under an equal-spp constraint, ellipsoidal path connection achieves the best performance, as it guarantees valid time-gated path samples.
However, when compared under equal-time budgets, our shrink mapping demonstrates an advantage, especially in complex scenes where ellipsoidal sampling incurs huge overhead due to scene–ellipsoid intersection tests.
We also show shrink mapping with spatial reuse (red curve), which can further improve performance, although it remains biased due to $R = 1$.

In summary, while shrink mapping does not guarantee exact path-length correctness in the same way as analytic ellipsoidal intersection, it offers a substantial performance advantage, particularly at $R = 1$, at the cost of bias, but without introducing blurring.


\subsection{Real-time Time-Gated Rendering with Temporal Reuse}
Path-length-aware reuse can also be applied to temporal reuse across frames.
Following~\citet{Lin:2022:Generalized}, we use a single iteration of spatial reuse with three neighbor selections and set $M_\text{cap} = 5$–$20$ (maximum confidence in \cref{eq:GRIS_MIS_weight}), while generating multiple initial paths per frame.
As in other ReSTIR-based methods, we exploit motion vector information to determine the corresponding pixel from which to fetch the reservoir in the previous frame.
For simplicity, we use direct connection for initial samples here, but it can be combined with ellipsoidal connection or shrink mapping described earlier.

\begin{figure*}
\includegraphics[width=\linewidth]{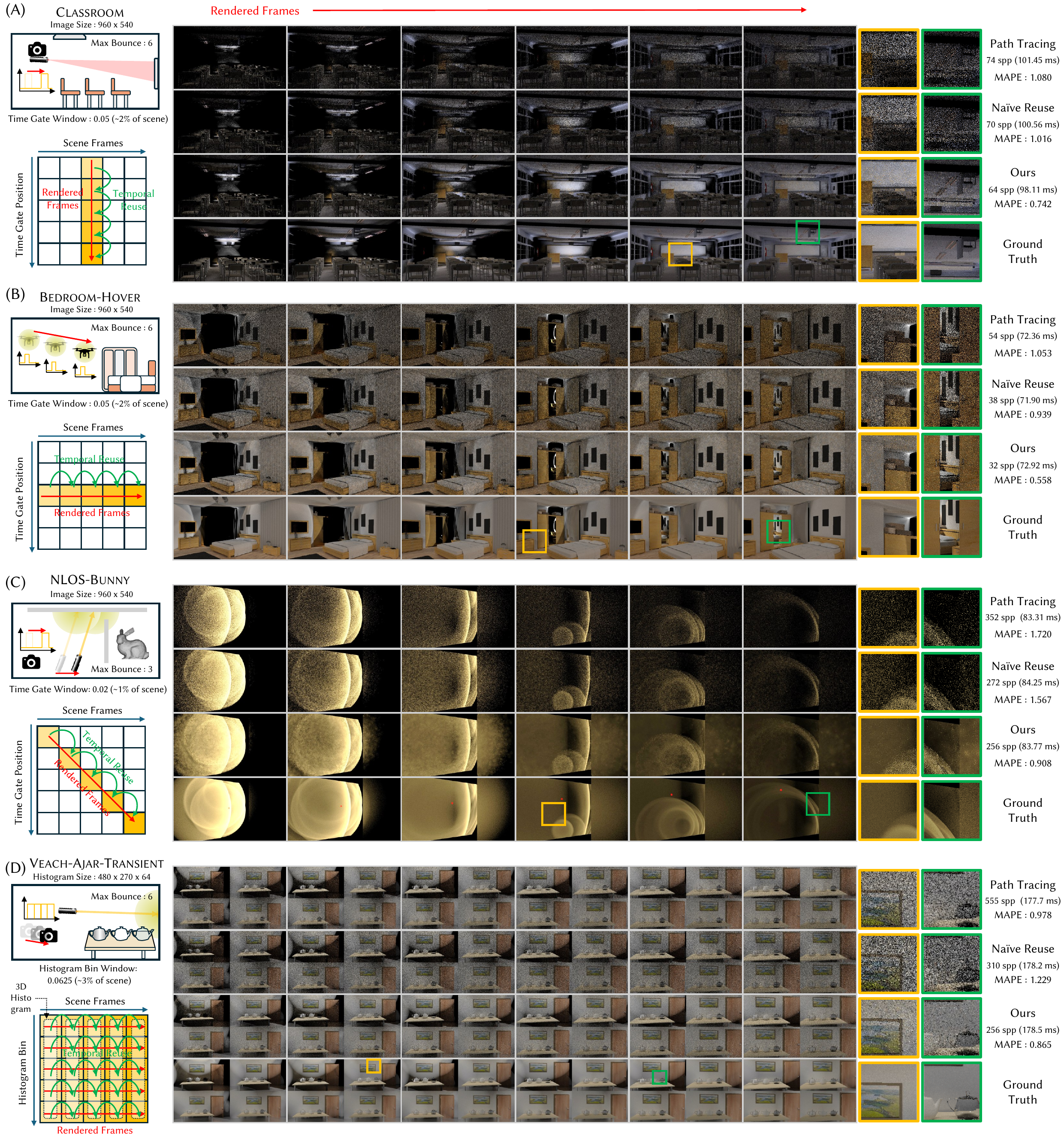}
\caption{Four scenarios demonstrating the proposed path temporal reuse.
(A) Streamed transient rendering in a static \textsc{Classroom} scene.
(B) Dynamic proximity camera simulation in a \textsc{Bedroom} scene.
(C) Transient NLOS imaging with a time-gated camera and a moving collimated laser; this setup is also used in later sections.
(D) Transient histogram rendering with the number of bins set to $B = 64$.
Across all scenarios, our method consistently produces the highest-quality results.
}
\label{figure:temporal_reuse_examples}
\end{figure*}

\paragraph{Real-time Transient Rendering}
We can exploit temporal reuse for real-time transient simulation in static scenes, producing transient images in a streaming fashion.
This is less common in the ToF community than in gaming or real-time rendering, but it is still useful for rapid inspection, as constructing a full histogram requires long accumulation times.
We visualize a real-time transient rendering result for the \textsc{Classroom} scene in~\cref{figure:temporal_reuse_examples}-(A).

\paragraph{Dynamic Scene}
We further demonstrate time-gated rendering in dynamic scenes, which is relevant for simulating proximity cameras or transient sensors that form image sequences by sweeping the time gate.
In~\cref{figure:temporal_reuse_examples}-(B), we simulate a proximity-sensing drone equipped with a fixed time-gated sensor and a collocated point light hovering within the \textsc{Bedroom} scene.
Our method produces substantially improved results compared to path tracing and naive reuse in this dynamic setting, and can even faithfully reproduce mirror-reflected regions.
We also evaluate the \textsc{NLOS-Bunny} scene in~\cref{figure:temporal_reuse_examples}-(C).
In this scene, the camera remains fixed while a collimated laser beam scans across the back wall, indicated by the red dot in the ground-truth image.
The time gate also shifts as the laser moves.
Again, our method yields superior results compared to baseline approaches.
We also demonstrate a downstream application for object shape reconstruction under a similar scene setting, which is described in detail in \cref{sec:nlos_reconstruction}. 


\begin{figure} 
\includegraphics[width=0.95\linewidth]{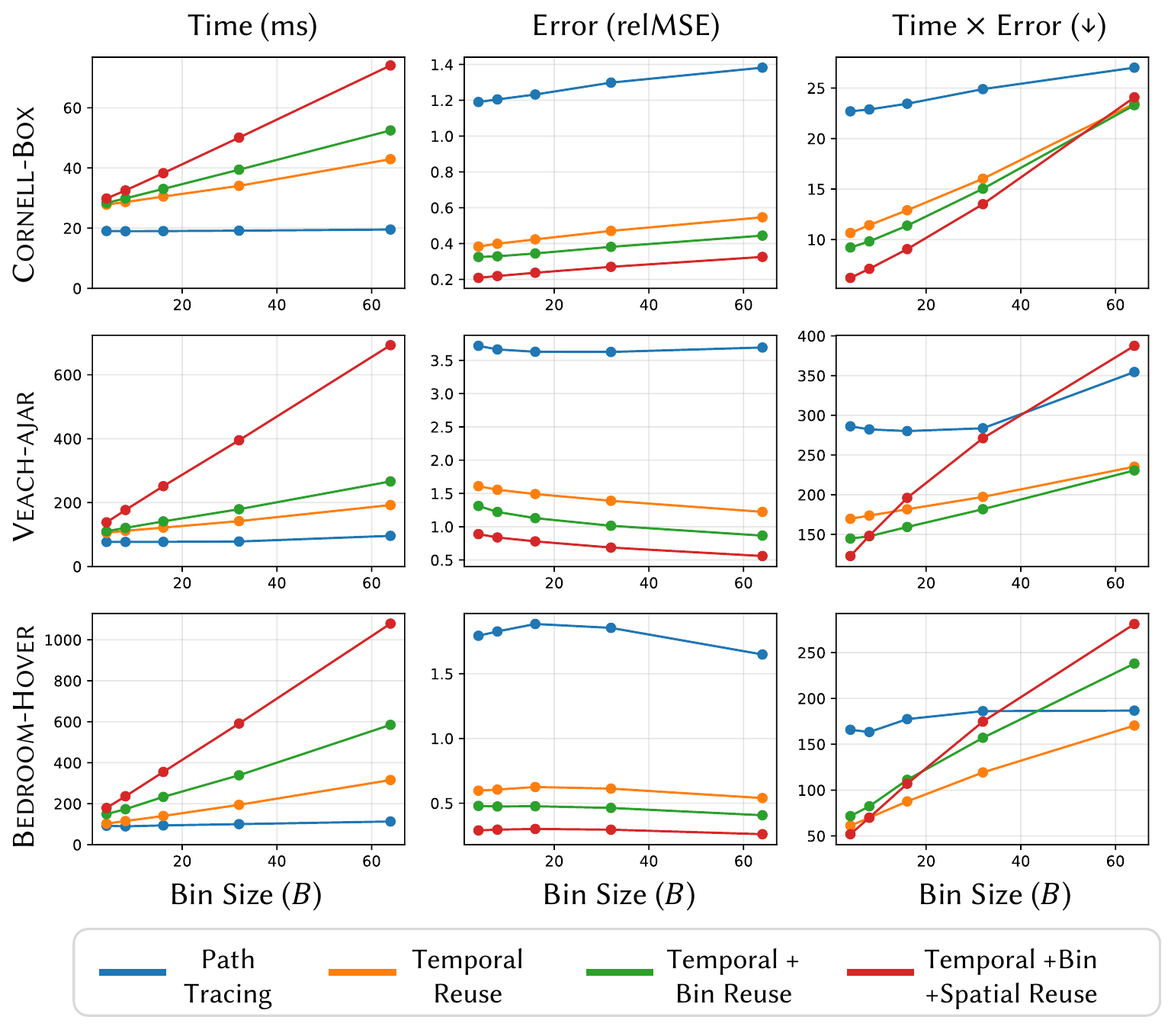}
\caption{Comparison of path reuse in transient histogram rendering with different $B$s. We could find that the cost-to-performance efficiency of path reuse diminishes as $B$ becomes larger.}
\label{figure:transient_histogram_evaluation}
\end{figure}

\subsection{Evaluation of Transient Histogram Rendering} 
We also evaluate our algorithm on transient histogram rendering, as discussed in~\cref{sec:transient_histogram_rendering}.
In this setting, the goal is to render a sequence of transient images simultaneously, producing an output histogram of size $(H, W, B)$ per frame.
An example of transient histogram rendering is shown in~\cref{figure:temporal_reuse_examples}-(D), where the camera moves forward in the \textsc{Veach-Ajar} scene.
We use a histogram bin count of $64$, and to limit reuse cost, we apply temporal reuse only.
We find that our method produces higher-quality results for transient histogram rendering compared to standard path tracing and naive reuse.

\paragraph{Cost-to-Performance Evaluation with Different $B$}
We evaluate the effectiveness of path reuse while accounting for its computational cost under different values of $B$.
Specifically, we measure rendering error, execution time, and their product as a simple indicator of cost-to-performance efficiency.
For the metric $\text{error} \times \text{time}$, lower values indicate better cost-to-performance.
Since second-order error decreases linearly over time, we use relative MSE as the error metric.
We evaluate several path reuse configurations, including temporal reuse, bin-wise reuse, and spatial reuse.
All experiments are conducted with $B \in [4, 64]$, while keeping the temporal width of each histogram bin fixed.
Each configuration is rendered for 100 frames with an animated camera, and the same spp is used across all experiments.

The results are shown in~\cref{figure:transient_histogram_evaluation}.
The first column reports rendering time.
As expected, the cost of naive path tracing remains nearly constant even $B$ increases, whereas the cost of path reuse grows linearly with $B$.
Configurations with more aggressive reuse exhibit steeper slopes due to higher reuse overhead.
The second column shows the relative MSE averaged over all bins.
The error remains nearly constant across different values of $B$, with minor variations depending on the histogram bin range.
Notably, more aggressive path reuse generally yields lower error.
The third column presents the cost-to-performance metric computed as time multiplied by error.
For small values of $B$, aggressive path reuse is clearly beneficial, as the reuse overhead is relatively small compared to the resulting error reduction.
As $B$ increases, the benefit of path reuse diminishes, and beyond a certain point, disabling path reuse becomes more advantageous.
We further observe that this trend is strongly influenced by the path reuse cost $C_\text{reuse}$.
The \textsc{Cornell-Box} scene exhibits the lowest reuse cost, while \textsc{Veach-Ajar} incurs a higher cost due to its textured materials.
The \textsc{Bedroom} scene shows the highest reuse cost, as the collocated moving light source requires updating the path throughput beyond the reconnection vertex.

In summary, increasing $B$ reduces the benefit of complex path reuse strategies, and in practice, a limited amount of reuse (e.g., temporal only) is often preferable.

\begin{figure}
\includegraphics[width=0.85\linewidth]{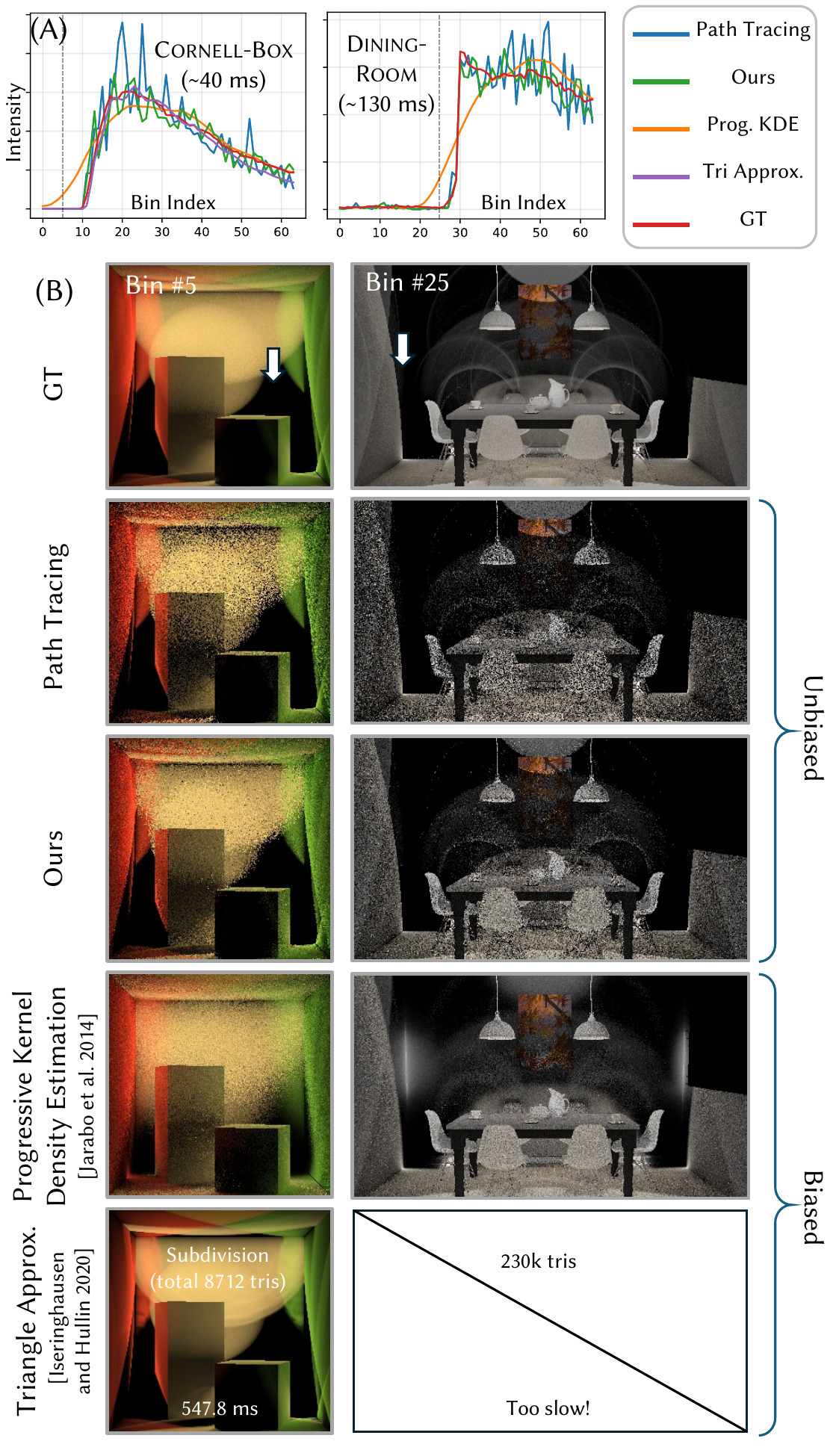}
\caption{Comparison of our method with progressive KDE~\cite{Jarabo:2014:Framework} and the linear triangle approximation~\cite{Iseringhausen:2020:Nonlineofsight}, both of which are biased.
Progressive KDE produces less noisy results but exhibits noticeable temporal blurring.
The triangle approximation works only for simple scenes and does not scale to complex geometry.
The transient profile at a specific point is shown in (A), and the full image is shown in (B).
}
\label{figure:transient_prev_comparison}
\end{figure}

\paragraph{Comparison with Biased Transient Rendering}
We compare our method with two previous transient rendering approaches: progressive kernel density estimation (KDE)~\cite{Jarabo:2014:Framework} and the radiosity-based linear triangle approximation~\cite{Iseringhausen:2020:Nonlineofsight}.
For progressive KDE, we use the Epanechnikov kernel with a kernel-width reduction rule $\Delta \tau_{j+1} / \Delta \tau_j = (j+\alpha)/(j+1)$, with $\alpha = 0.8$, following \citet{Jarabo:2014:Framework}.
For the triangle approximation, we subdivide the scene geometry to obtain a sufficiently fine mesh (8,712 triangles for \textsc{Cornell-Box}).
For our method, we apply temporal reuse only.
All methods are evaluated over 100 frames using $B = 64$ and the same time budgets, except for the triangle approximation.
We visualize the transient profile at a specific point and the full image in~\cref{figure:transient_prev_comparison}, at a specific frame.
Compared to unbiased methods, progressive KDE produces less noisy images at the cost of introducing bias, primarily in the form of temporal blurring.
As shown in~\cref{figure:transient_prev_comparison}-(A), this bias manifests as non-zero values in regions where the ground truth is zero.
The triangle approximation requires heavy tessellation and takes 547.8 ms even for the simple \textsc{Cornell-Box} scene, making it impractical for more complex scenes such as \textsc{Dining-Room}, which contains approximately 230k triangles.
Moreover, it does not support multi-bounce light transport.

Our method is unbiased and produces higher-quality results than naive path tracing under the same time budget.
While the results remain noisier than those of biased methods, this gap highlights opportunities for future improvement.



\subsection{Evaluation of Doppler Frequency Rendering}
\begin{figure}
\includegraphics[width=\linewidth]{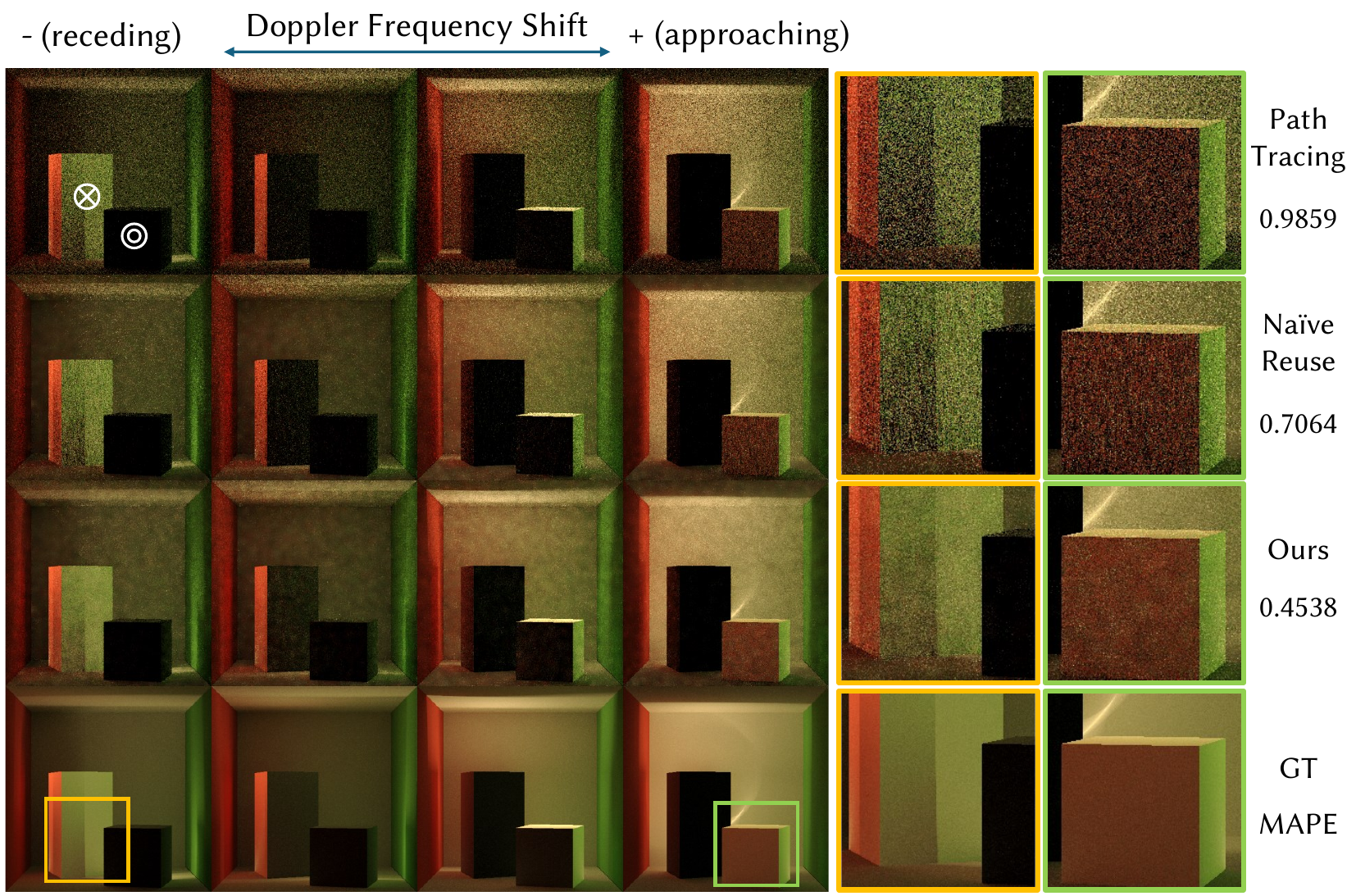}
\caption{Doppler frequency shift rendering under different frequencies. The large box is receding from the camera, resulting in a negative frequency shift, while the small box is approaching the camera, producing a positive frequency shift. Overall, our proposed algorithm also performs well under path-velocity constraints.}
\label{figure:doppler_result}
\end{figure}

Finally, we demonstrate a simple example of Doppler frequency rendering in~\cref{figure:doppler_result}.
In this scene, the large box is receding from the camera, while the small box is approaching it.
Each image is rendered with a 2-second budget in an offline style, with only spatial reuse performed.
Across different Doppler frequency shifts, our method produces improved results both qualitatively and quantitatively, exhibiting performance gains similar to those observed for the path-length constraint.

%% file: sections/07_applications_v1.0.tex
\section{Downstream Applications}
Our renderer can be easily deployed to test downstream applications. This capability has the potential to reduce the replication crisis in computational imaging while also accelerating the software development of imaging systems. Below, we demonstrate two applications of the renderer for downstream tasks. 


\begin{figure}
\includegraphics[width=0.90\linewidth]{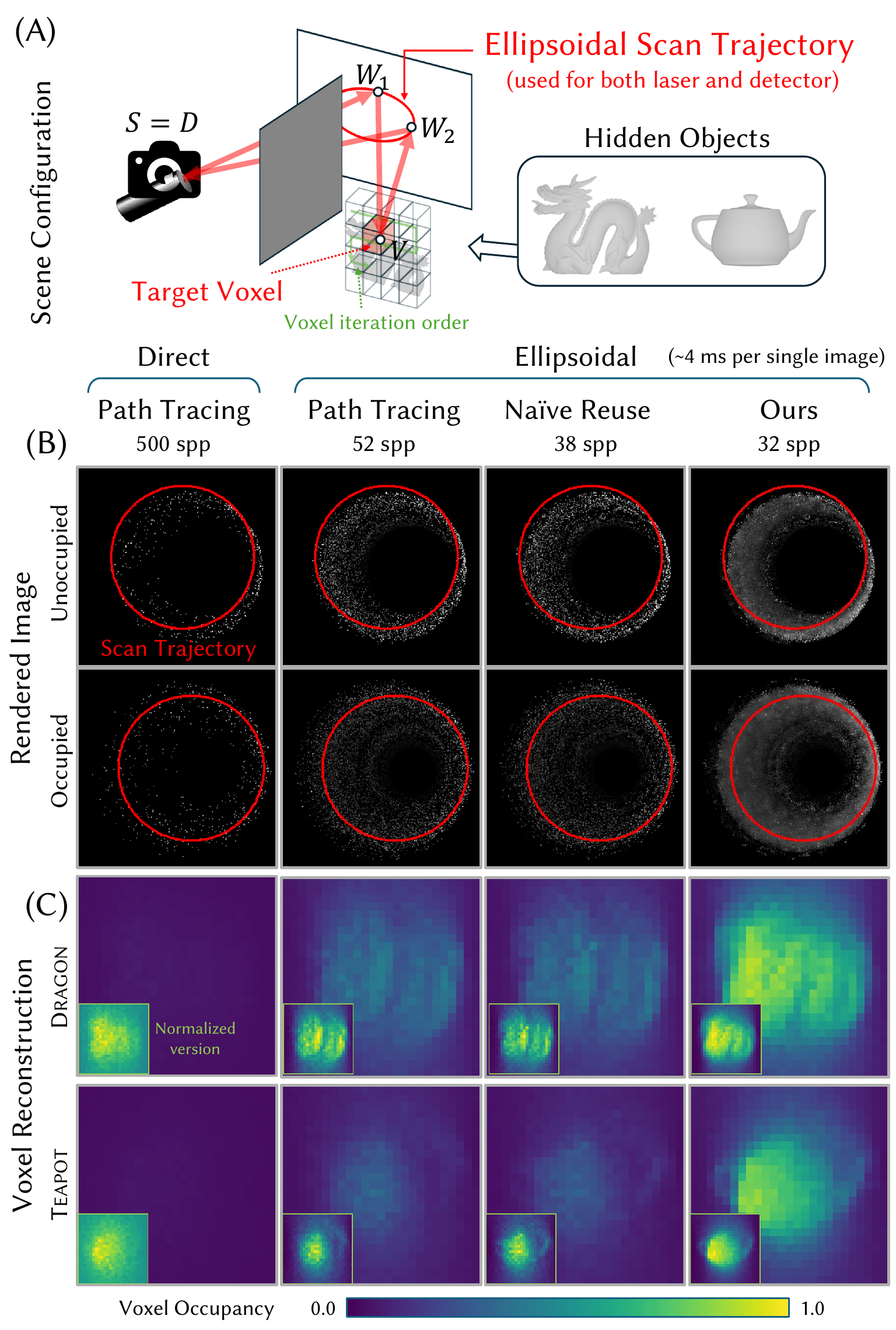}
\caption{Simulation of NLOS imaging with temporal focusing~\cite{Pediredla:2019:SNLOS} using our rendering pipeline.
(A) Scene configuration for detecting a hidden object behind a wall. Both the laser and the sensor scan ellipsoidal trajectories on the visible wall to determine whether a target voxel is occupied.
(B) Example rendered image, and (C) reconstructed voxel grids (the version normalized by the maximum value is shown in the bottom-left), obtained using different methods under a time budget of 4~ms per image, demonstrating the advantage of our proposed method.}
\label{figure:snlos}
\end{figure}

\subsection{NLOS Reconstruction with Temporal Focusing}
\label{sec:nlos_reconstruction}
We first simulate NLOS detection with temporal focusing, similar to \citet{Pediredla:2019:SNLOS}, but with a small change to allow for better signal processing. The scene configuration is illustrated in \cref{figure:snlos}–(A). In this setup, the target object is hidden behind an occluder and is not directly visible to the imaging system, which can directly see only a diffuse relay wall. The objective is to recover the voxel reflectance (albedo) of the hidden object by exploiting indirect multi-bounce reflections and ToF measurements.
We will first describe the imaging setup and then the rendering strategy. 

\paragraph{Imaging setup}
In \citet{Pediredla:2019:SNLOS}'s setup, the imaging system consists of a single-pixel time-gated sensor ($D$) and a laser source ($S$) that are colocated. Instead of a single-pixel detector, we assume an array time-gated sensor such as an ICCD.
The key idea of \citet{Pediredla:2019:SNLOS} is to temporally focus the illumination and backprojected imaging rays onto a single voxel $V$ and image the NLOS scene by scanning one voxel at a time.

First, consider all light paths that temporally focus on a single voxel $V$ at time gate $\tau_1$. Light emitted from the laser first reflects off the visible relay wall and then arrives at $V$. Therefore, points $W_1$ on the relay wall that lead to temporal focusing at $V$ satisfy
\begin{equation}
\len{S - W_1} + \len{W_1 - V} = \tau_1.
\label{eq:tempfocus1}
\end{equation}
Similarly, light reflected from $V$ will focus on the detector at time $\tau_2$ after reflecting off wall points $W_2$ that satisfy
\begin{equation}
\len{V - W_2} + \len{W_2 - D} = \tau_2.
\label{eq:tempfocus2}
\end{equation}

Both \cref{eq:tempfocus1} and \cref{eq:tempfocus2} are equations of an ellipse; hence, by illuminating and imaging an ellipse on the relay wall, we can compute the voxel reflectance.
In \citet{Pediredla:2019:SNLOS}, $\tau_1 = \tau_2 = \tau$.

This technique only sums imaging rays along the ellipse. Instead, by using an array detector, we can treat all points along the imaging ellipse independently and develop improved signal-processing techniques for accurate NLOS scene estimation. Specifically, we exploit the observation that the entire imaging ellipse must be visible at the specific time gate $\tau$, and thus we take the geometric mean of the ellipsoidal points. The geometric mean yields a better signal-to-background ratio than the arithmetic mean.

\paragraph{Rendering setup}
To simulate this process using our renderer, we render multiple images while steering the laser toward different positions along the ellipsoidal trajectory.
In each rendered image, we take the geometric mean along the imaging ellipse to estimate the voxel reflectance.
The target time range is set to $2\tau$, and $\tau$ is chosen to be slightly larger than $\len{S - C} + \len{C - V}$, where $C$ denotes the center of the visible wall.
An example of the rendered image and the corresponding sensor ellipsoidal trajectory for unoccupied and occupied voxels is shown in~\cref{figure:snlos}–(B).

We compare four different methods: path tracing with direct connection, path tracing with ellipsoidal connection, ellipsoidal connection with naive path reuse, and ellipsoidal connection with our proposed method.
All images are rendered at a resolution of $256 \times 256$ under a time budget of 4~ms per image.
The laser is scanned over 16 positions, and we scan $25 \times 25$ NLOS voxels, resulting in a total of $16 \times 25 \times 25$ rendered images.
The time-gate width is set to $0.01$, corresponding to less than $1\%$ of the scene scale.
We apply both temporal and spatial reuse for reuse-based methods.
To ensure frame-to-frame similarity between rendered images (which makes temporal reuse useful), we scan voxels in a continuous order, as indicated by the bright green trajectory in \cref{figure:snlos}–(A).

The voxel reconstruction results are shown in~\cref{figure:snlos}–(C).
As observed by~\citet{Pediredla:2019:SNLOS}, direct connection performs poorly in this scenario because most rays do not contribute to the target voxels.
Ellipsoidal path connection improves reconstruction quality, but the results remain noisy.
Naive path reuse provides some improvement, such as partially recovering the teapot spout.
In contrast, our method produces the most accurate reconstruction, owing to the significantly reduced noise in the rendered images, as shown in~\cref{figure:snlos}–(B).
We use a simple scene configuration here, but more challenging scenarios, such as scenes with multiple objects, non-diffuse materials, and higher-order bounces, could be explored in future evaluations.

\begin{figure} 
\includegraphics[width=\linewidth]{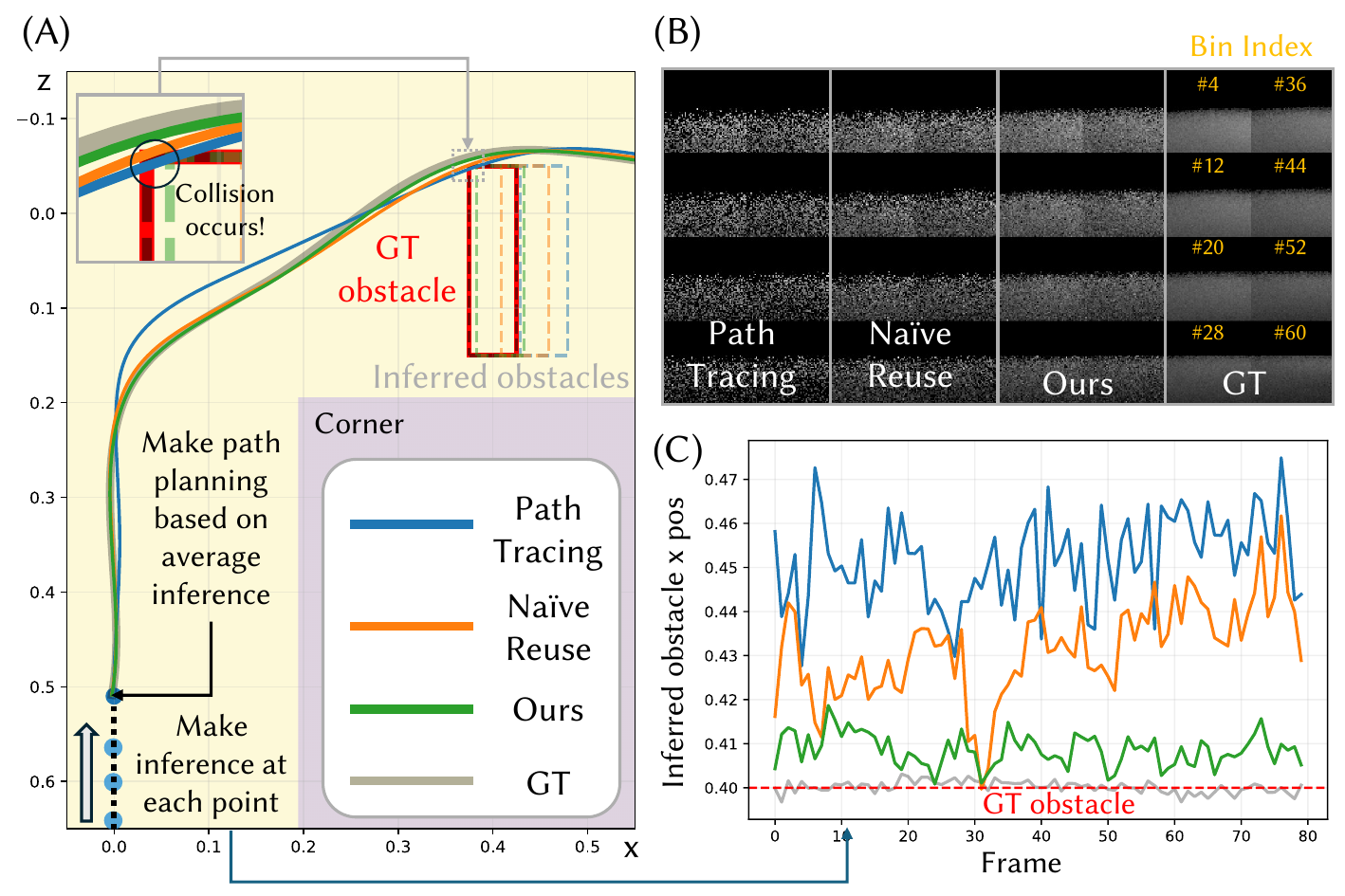}
\caption{
We simulate NLOS navigation~\cite{young2025enhancing} using our rendering pipeline.
(A) Overall scene configuration, where a camera with a collocated light source moves forward and captures transient histograms.
(B) Rendered transient image used for inference.
(C) Inferred hidden object position.
When the camera reaches $z = 0.5$, path planning is performed based on the averaged object inference.
Compared to path tracing and naive reuse, which result in collisions with the true obstacle, our method achieves more accurate object localization and consequently more robust path planning.
}
\label{figure:nlos_navigation}
\end{figure}


\subsection{NLOS Navigation}
We also simulate an inference-based NLOS navigation scenario similar to~\citet{young2025enhancing}. We consider an L-shaped road with an obstacle located in an occluded region. The goal is to detect the hidden object from global transient illumination and enable safe, fast, and robust path planning. The overall scene configuration is illustrated in~\cref{figure:nlos_navigation}–(A).

The camera is initially positioned at $(x, z) = (0, 1)$ and moves toward the $(0, -1)$ direction, starting from a location outside the field of view. As it advances, the system continuously infers the location of the hidden object and, at a predefined point ($z = 0.5$), performs path planning based on the accumulated estimates. For simplicity, we infer only the $x$-coordinate of the object, assuming that the $z$-coordinate and object size are known. The camera and laser are collocated, with the laser having a small opening angle and a slight downward tilt toward the ground.

We train a lightweight convolutional neural network (CNN) for object location inference. The network takes as input the camera position, viewing direction, and transient histogram data, and outputs the estimated $x$-coordinate of the object. Training is performed using ground-truth rendered data with randomly sampled object positions, while inference uses noisy data generated by different rendering methods. An equal time budget of 10~ms per frame is enforced for all methods. We use temporal reuse only. The per-frame histogram resolution is $64 \times 64$ with $B = 64$. The CNN consists of two convolutional layers followed by a fully connected layer.

The rendered transient histograms and the corresponding inference results are shown in~\cref{figure:nlos_navigation}–(B) and (C), respectively. In this experiment, the object is located at $x = 0.4$. Path tracing and naive reuse produce inaccurate estimates due to noise in the rendered histograms. In contrast, our method yields more accurate inference results by reducing noise through temporal path reuse.

The accuracy of object inference directly affects path planning quality. We visualize the ground-truth obstacle position and the planned paths produced by each method using the A* algorithm with spline smoothing. Because path tracing and naive reuse predict the object to be much farther than its true position, the resulting paths collide with the actual obstacle. In contrast, our method produces a more accurate estimate of the object location, enabling safer and more robust navigation. While we focus on a simple scenario here, more advanced navigation strategies and more complex environments could be explored using our simulation framework in future work.

%% file: sections/08_conclusion_v1.1.tex
\section{Discussion and Conclusion}
In this manuscript, we introduce a path-length-aware shift mapping framework that enables effective spatio-temporal sample reuse for time-of-flight (ToF) rendering.
By explicitly enforcing optical path-length constraints during reuse, our method overcomes a key limitation of steady-state ReSTIR when applied to time-resolved imaging, where naive reuse often produces temporally invalid contributions.
The proposed approach supports a wide range of ToF rendering scenarios, including offline time-gated rendering, interactive transient rendering, transient histogram rendering in dynamic scenes, and extensions to Doppler frequency simulation.
We further demonstrate two downstream applications---shape reconstruction and navigation---highlighting the potential of our renderer for interactive and decision-driven ToF sensing systems.

\paragraph{Limitations and Future Works}
We discuss the limitations of our approach and outline potential directions for future research.

First, we use the reconnection vertex in the ReSTIR-PT framework as the adjustment vertex for path-length-aware shift mapping.
More sophisticated heuristics for selecting this vertex---such as incorporating geometric or material properties of adjacent vertices $\pbf_1$ and $\pbf_2$---could improve both robustness and efficiency.
Alternatively, instead of perturbing a single vertex, one could distribute the path-length adjustment across the entire path to minimize per-segment changes, although this would incur additional cost due to updating the full path.
Moreover, our method primarily enforces path-length constraints and does not explicitly account for specular manifolds.
In scenes where both temporal and specular constraints are important, a joint manifold formulation would be required, and designing efficient strategies for such mixed constraints remains an open challenge.


We currently employ a basic Newton-based solver for manifold traversal, but more advanced variants from steady-state caustics rendering~\cite{fan2024specular, hong2025sample} could be adapted to improve convergence and stability.
Also, while we select the update direction based on the gradient to minimize vertex displacement, this heuristic does not always guarantee valid reuse due to visibility changes (e.g., the discrepancy between Newton success rate and actual reuse success rate in~\cref{tab:newton_statistics}).
Developing more robust direction-selection strategies that incorporate visibility and geometric constraints is an important direction for future work.

Our current implementation also assumes delta light sources, which precludes searching over extended light source domains. 
For finite-area light sources, the constraint simplifies to the distance between the previous vertex and the light source, allowing sampling directly over the emitter surface. 
Also in this setting, spherical connections are more appropriate than ellipsoidal ones, as discussed by~\citet{Pediredla:2019:Ellipsoidal}.

Finally, there is substantial room for improvement in transient histogram rendering. In this work, we directly apply path-length-aware shift mapping to each histogram bin.
However, since transient histograms involve continuous time bins rather than strict delta manifolds, explicitly enforcing path-length constraints via root finding may be unnecessary. 
Time-bin splatting strategies, similar to those proposed by~\citet{liu2025reservoir}, could provide a more efficient alternative. 
A remaining challenge is the design of effective backup sampling strategies for time bins that receive no prior samples. 
Developing more scalable ReSTIR-based algorithms for large numbers of time bins $B$ is also an important direction for future work.

\begin{acks}
We acknowledge the support of the Visual Computing Seminar and colleagues from the RISC and VCL labs for their helpful feedback on this project.
This work was supported in part by the National Science Foundation under Awards 2403122 and 2326904.
\end{acks}